\documentclass[acmsmall]{acmart}
\usepackage{url}
\usepackage{graphicx}
\usepackage{amsmath}
\usepackage[ruled, linesnumbered, noend]{algorithm2e}
\usepackage{algpseudocode}
\usepackage{tabularx}
\usepackage{graphicx}
\usepackage{pdflscape}
\usepackage{rotating}
\usepackage{multirow}
\usepackage{hyperref}
\usepackage{natbib}

\setcopyright{acmlicensed}
\copyrightyear{2025}
\acmYear{2025}
\acmDOI{XXXXXXX.XXXXXXX}
\acmISBN{978-1-4503-XXXX-X/2018/06}

\begin{document}

\title{VulRG: Multi-Level Explainable Vulnerability Patch Ranking for Complex Systems Using Graphs}

\author{Yuning Jiang}
\email{yuning\_j@nus.edu.sg}
\orcid{0000-0003-4791-8452}
\affiliation{%
  \institution{National University of Singapore}
  \country{Singapore}
}
  
\author{Nay Oo}
\email{nay.oo@ncs.com.sg}
\affiliation{%
 \institution{NCS Cyber Special Ops R\&D}
  \country{Singapore}}

  \author{Qiaoran Meng}
\email{qiaoran@nus.edu.sg}
\affiliation{%
  \institution{National University of Singapore}
  \country{Singapore}}

\author{Hoon Wei Lim}
\email{hoonwei.lim@ncs.com.sg}
\affiliation{%
 \institution{NCS Cyber Special Ops R\&D}
  \country{Singapore}}

\author{Biplab Sikdar}
\email{bsikdar@nus.edu.sg}
\affiliation{%
  \institution{National University of Singapore}
  \country{Singapore}}

\renewcommand{\shortauthors}{Jiang et al.}

\begin{abstract}
As interconnected systems proliferate, safeguarding complex infrastructures against an escalating array of cyber threats has become an urgent challenge. The growing number of vulnerabilities, coupled with resource constraints, makes addressing every vulnerability impractical, thereby rendering effective prioritization essential. However, current risk prioritization methods, which often rely on expert judgment or focus solely on exploit likelihood and consequences, lack the granularity and adaptability needed for complex systems. This paper presents a novel graph-based framework for patch prioritization that optimizes security in complex systems by integrating diverse data sources and metrics into a universally applicable model. We introduce refined risk metrics that enable detailed assessments at the component, asset, and system levels. The framework employs two key graphs: a network communication graph to model potential attack paths and identify the shortest routes to critical assets, and a system dependency graph to capture risk propagation from exploited vulnerabilities across interconnected components. By defining asset criticality and component dependency rules, our approach systematically assesses and mitigates risks. Benchmarking against state-of-the-art methods demonstrates our framework’s superior accuracy in vulnerability patch ranking, with enhanced explainability. This framework not only advances vulnerability management but also sets the stage for future research in adaptive cybersecurity strategies.
\end{abstract}



\maketitle

\section{Introduction}

In our increasingly interconnected and digitized world, securing complex systems is vital for protecting sensitive information and ensuring the continuity of critical operations. The growing number of vulnerabilities within these systems necessitates a strategic approach to remediation. Given the sheer volume of vulnerabilities and the reality of limited resources, it is impractical to address every threat, making the prioritization of defensive measures crucial \cite{iannone2022secret} \cite{le2022survey} \cite{mehri2023automated}.

Existing studies in vulnerability management provide valuable insights into statistical prioritization techniques. However, they often fail to adapt to rapidly evolving threats and overlook contextual dependencies among system vulnerabilities \cite{he2019unknown}. These methods typically rely on static data sources, expert-based judgments, or simplistic risk scores, such as those from the Common Vulnerability Scoring System (CVSS), leading to sub-optimal prioritization outcomes \cite{howland2023cvss} \cite{esposito2023can}. Some approaches rank vulnerabilities based solely on exploit likelihood \cite{jacobs2021exploit} \cite{alperin2019risk} or consequence, often due to limited data availability \cite{angelelli2024robust} \cite{de2023no}. However, these methods generally do not account for the dynamic nature of system architecture and its inter-dependencies \cite{zeng2023illation} \cite{imtiaz2021comparative}, limiting their applicability in rapidly evolving threat environments and lack a standardized methodology for integrating diverse data sources and metrics into a universally applicable framework. This gap hinders their effectiveness in diverse organizational settings.

\begin{figure*}[h]
\centering
\includegraphics[width=1\textwidth]{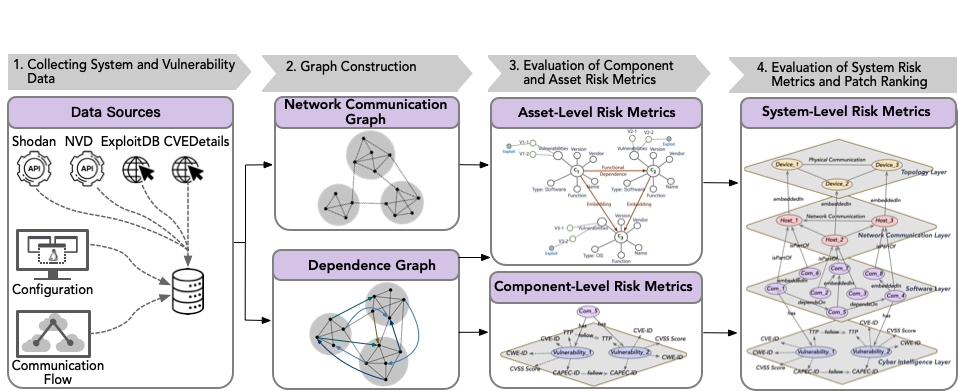}
\caption{The framework of the graph-based risk aggregation approach.}
\label{fig:Architecture}
\end{figure*}

To address these limitations, this paper introduces VulRG, a graph-based multi-level patch-prioritization framework specifically tailored to the complexities of modern systems, as illustrated in Fig. \ref{fig:Architecture}. Our algorithm not only assesses individual vulnerability risks but also incorporates an understanding of cumulative risk exposure, factoring in component dependencies, exploit likelihood, impact, and risk propagation, to provide quantitative risk assessment for component, asset and system levels \cite{lei2022hybrid}. By integrating commonly used and open-source vulnerability data sources such as National Vulnerability Database (NVD), Exploit Database \textit{ExploitDB}, and \textit{CVEdetails}, we retrieve vulnerability attributes such as exploits, utilized Ransomware, CVSS assessment and Exploit Prediction Scoring System (EPSS) score. Our framework introduces two critical graph models: the network communication graph and the system dependency graph. The network communication graph models potential attack paths, identifying the shortest routes attackers might use to reach critical assets. The system dependency graph captures how exploited vulnerabilities can propagate risks across interconnected components, affecting the overall system \cite{denuit2022conditional}. To enhance the precision of our risk assessments, we define clear criticality levels for assets, ensuring that the most vital system components receive appropriate attention. Additionally, we establish dependency rules for components, which further refine the risk propagation analysis within the dependency graph. This holistic approach ensures a more accurate and strategic allocation of security resources, effectively protecting high-criticality assets. 

We have benchmarked our approach against state-of-the-art methods validated in real-world scenarios, demonstrating its practicality and significant impact on system security. Our analysis highlights the potential of our framework to enhance security across diverse, real-world environments. 


\subsection{Problem Statement}

\begin{figure}[h]
\centering
\includegraphics[width=0.4\textwidth]{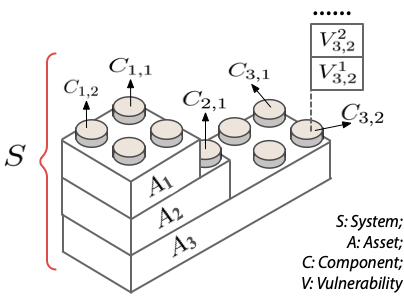}
\caption{Connections between system, asset and component.}
\label{fig:SystemAssetCom}
\end{figure}

Consider $S$, a complex system comprised of $N$ components denoted as $C_i$. Each component $C_i$ is associated with $V_i$ vulnerabilities, where $V_i$ indicates the number of vulnerabilities per component. Our goal is to develop an algorithm $P$ that assigns a system-level risk score $\mathbf{R}_{\text{system}}$ that is aggregated from risk scores of the components, as shown in Fig. \ref{fig:SystemAssetCom}. To enhance the overall security posture of the system, we aim to identify and prioritize the vulnerabilities whose remediation would result in the maximum reduction of the system-level risk score. We define $\mathbf{R}_{\text{system}}^{ij}$ as the system-level risk score after patching the $j$-th vulnerability in component $C_i$. The optimization objective is to maximize the reduction in the system-level risk by selecting the optimal vulnerability to patch, formulated as:

\[
\text{Maximize} \quad \max_i (\mathbf{R}_{\text{system}} - \mathbf{R}_{\text{system}}^{ij})
\]

In this paper, complexity pertains to the structural intricacy of the system being analyzed, including the number and diversity of assets, the interdependencies between them, and the variety of vulnerabilities present. Meanwhile, vulnerability mitigation needs to take into consideration of the budget and resource constrains, necessitating a holistic model to accurately assess risk and prioritize patches.

\subsection{Contributions}

This paper makes several significant contributions to the field of vulnerability management within complex systems:

\begin{enumerate}
\item \textbf{Dynamic Multi-Level Risk Aggregation:} We present a graph-based risk aggregation framework that adapts to changes in system dependencies and threats, allowing for risk assessment at multiple levels - component, asset, and system. This ensures accurate evaluation of both direct and propagated risks.

\item \textbf{Context-Aware Patch Ranking:} Our framework includes a patch ranking mechanism that prioritizes remediation based on vulnerability severity, exploit likelihood, component centrality, and system interdependencies. This ensures that patching efforts are aligned with the broader system impact. 
\end{enumerate}

The paper is structured as follows: Section \ref{sec:Literature} reviews the state-of-the-art in vulnerability prioritization and identifies key research gaps. Section \ref{sec:Theory} defines the theories and metrics employed for risk aggregation. In Section \ref{sec:Approach}, we introduce the proposed graph-based risk aggregation and vulnerability ranking method in a step by step manner. Section \ref{sec:CaseStudy} applies this approach to an enterprise network, generating multi-level vulnerability rankings to inform cybersecurity decisions. Section \ref{sec:Experiment} validates the effectiveness of our method by implementing it in scenarios drawn from recent literature, showcasing its accuracy and explainability. Finally, Sections \ref{sec:Discussion} and \ref{sec:Conclusion} offer discussions and conclusions, respectively.

\section{Background} \label{sec:Literature}

\subsection{\textit{CVSS} and \textit{EPSS}} \label{sec:CVSS}

A crucial step in vulnerability prioritization is the quantitative estimation of risk, defined by \textit{Probability} and \textit{Impact}  \cite{zeng2023illation}. \textit{Probability} measures the exploit likelihood of a vulnerability \cite{elder2024survey} while \textit{Impact} measures the consequence of that successful exploit against the vulnerable asset.

CVSS is developed and maintained by the Forum of Incident Response and Security Teams (FIRST) \cite{cvss}. As the leading method for vulnerability prioritization, it evaluates vulnerabilities by considering both exploitability and impact through its Base Metrics, which aggregate values derived from expert-defined measurements. Taking \textit{CVSS} version 3 (V3) as an example, it categorizes the principal metrics of vulnerabilities into three groups: \textit{Base}, \textit{Temporal}, and \textit{Environmental} metrics. The \textit{Base} metrics are further divided into \textit{Exploitability}, \textit{Scope}, and \textit{Impact} categories. The \textit{Exploitability} metrics comprise of \textit{Attack Vector (AV)}, \textit{Attack Complexity (AC)}, \textit{Privileges Required (PR)}, and \textit{User Interaction (UI)}. \textit{Scope} measures how a vulnerability in one component can affect other components, while \textit{Impact} metrics use \textit{Confidentiality (C)}, \textit{Integrity (I)} and \textit{Availability (A)}. \textit{Temporal} metrics reflect dynamic aspects of a vulnerability, such as changes in exploit techniques and patch availability. \textit{Environmental} metrics depend on deployment contexts. \textit{CVSS} combines these metrics to generate a score for a vulnerability, which helps in determining its severity. Then, \textit{CVSS} v3 rates vulnerabilities as \textit{'none'} (0.0), \textit{'low'} (0.1-3.9), \textit{'medium'} (4.0-6.9), \textit{'high'} (7.0-8.9), or \textit{'critical'} (9.0-10.0).

EPSS is a statistical model that estimates the likelihood of a vulnerability being exploited in the wild within the next 30 days \cite{jacobs2021exploit}. It provides a score between 0 and 1, where a higher score indicates a higher likelihood of exploitation. This score is predicted using a logistic regression model that evaluates features like software vendor, availability of exploit code, vulnerability characteristics (e.g., code execution, privilege escalation, etc.), and the number of references associated with a vulnerability. FIRST team provides EPSS API \cite{epss} that supports easy fetching of exploit-probability scores for vulnerabilities. 

\subsection{CVSS-Based Prioritization}

Many studies leverage CVSS as the foundation for vulnerability prioritization due to its ease of use \cite{walkowski2021vulnerability} \cite{angelelli2024robust} \cite{pecl2021manager}. For example, \cite{angelelli2024robust} utilized CVSS within a statistical framework using mid-quantile regression, introducing the Agreement of Grounded Rankings to maintain prioritization accuracy even with incomplete information. 

Despite its widespread use, critiques have emerged highlighting that CVSS alone may not adequately capture the likelihood and impact of a cyber attack \cite{howland2023cvss} \cite{esposito2023can}. To overcome the limitations of CVSS, several approaches have integrated external metrics, such as exploitability, threat intelligence, and environmental factors, to enhance the accuracy of vulnerability prioritization \cite{forain2022revs} \cite{wang2022automotive} \cite{northern2021vercasm}. For instance, \cite{walkowski2021automatic} enhanced CVSS base scores with environmental metrics such as Collateral Damage Potential (CDP) and Target Distribution (TD), to tailor prioritization to the characteristics of specific IT resources. Building on this, \cite{walkowski2021vulnerability} further automated the CVSS environmental scoring process, streamlining the workflow.

\subsection{Prioritization Integrating Exploit Metrics}

Efforts to improve patch prioritization often emphasize vulnerabilities with a higher probability of exploitation. For instance, \cite{jacobs2023enhancing} developed EPSS to predict the likelihood of exploitation within 30 days of a vulnerability's disclosure, leveraging over 1,400 features, including social media activity and public exploit code availability. Similarly, \cite{figueiredo2022exploitability} utilized genetically tuned neural networks to predict exploitation likelihood using open source vulnerability data sources like NVD and Exploit database. 

Building on this, \cite{yoon2023vulnerability} prioritized vulnerabilities based on an exploitation risk score that combines exploit chaining risk, exploit code availability, and exploit usage probability. Their system integrates data from sources like the MITRE ATT\&CK framework and CVE databases, making it particularly useful for vulnerability assessment in Operational Technology (OT) and Industrial Control System (ICS) environments, with a focus on Industrial Internet of Things (IIoT) devices.

Incorporating network-specific information and probabilistic models into vulnerability prioritization is another trend in this area. \cite{zeng2021licality} and \cite{zeng2023illation} proposed the LICALITY framework and its upgrade approach ILLATION, integrating neural networks and probabilistic logic programming for threat modeling, prioritizing vulnerabilities based on exploit likelihood and network configurations.

\subsection{Context-Aware Prioritization}

Context-aware models in vulnerability prioritization take a broader system perspective by incorporating factors such as network topology, attack paths, and business impact.

Network-specific models and attack path analysis play a significant role in context-aware prioritization. For instance, \cite{wang2023critical} utilized a Knowledge Graph (KG) approach to identify exploitable vulnerabilities in ICS, prioritizing them based on relationships between vulnerabilities, assets, and attack scenarios. Attack graphs and Bayesian networks are also commonly employed to map relationships between vulnerabilities and potential attack paths, providing insights into how vulnerabilities could be exploited. For example, \cite{homer2013aggregating} combined attack graphs with probabilistic reasoning to quantitatively assess network security, aggregating component metrics from CVSS vectors into likelihood estimates of an attacker gaining network privileges. Similarly, \cite{yan2022cyber} used attack-path graphs to evaluate network security vulnerabilities in SCADA systems, assessing the impact of cyberattacks on power systems through a minimum shedding load model.

Graph-based models have proven effective in representing system dependencies. For instance, \cite{longueira2022novel} introduced a dependency graph model to track vulnerabilities and prioritize patches by representing the internal structure of industrial components. Building on this, \cite{longueira2022gotta} modeled system dependencies, aggregating CVSS scores and considering contextual factors like functionality disruption and deployment context.

Business impact is another key factor in context-aware prioritization \cite{keskin2021scoring}. \cite{keskin2021scoring} assessed the functional dependencies between assets and business processes, analyzing how vulnerabilities in one asset could affect others. Likewise, \cite{ahmadi2022automated} developed a framework for automating vulnerability risk management by integrating data from multiple sources and considering organizational context factors such as asset exposure and business impact. Incorporating organizational goals into prioritization has also been emphasized. 

Some studies have focused on adapting CVSS to specific domains by incorporating additional metrics. For instance, \cite{wang2022automotive} modified the standard CVSS framework by integrating metrics tailored to automotive cybersecurity, including safety, finance, operation, and privacy. Similarly, \cite{jung2022cavp} developed a system that automatically derives temporal metrics for vulnerabilities using expert-validated heuristic rules, while also considering CVSS environmental metrics.

In terms of security metrics aggregation, \cite{ahmed2018aggregation} proposed a reference architecture that aggregates security metrics through data collection, normalization, and visualization via dashboards, providing decision-makers with insights into security states at various levels of granularity.

\subsection{Research Gap}

While significant progress has been made in vulnerability prioritization and risk assessment, a comprehensive framework that integrates multi-dimensional metrics and considers both static and dynamic cyber elements is still lacking \cite{fumagalli2023semantics}. Current models tend to focus on specific aspects of vulnerability management, often missing the holistic perspective needed to address the complexity of interconnected systems. Bridging this gap requires a shift towards more integrated and adaptable models that can apply to various system configurations. 

This work presents a graph-based patch prioritization framework that addresses these limitations by incorporating dynamic risk assessments and modeling cumulative risk exposure across interconnected components. The proposed framework leverages network communication and system dependency graphs to model potential attack paths and risk propagation, enabling more accurate vulnerability prioritization.

\section{Theoretical Definitions} \label{sec:Theory}

\subsection{System, Host, Asset and Component}

Our definitions of system, host, asset, and component are built upon existing standards such as ISO/IEC 27001:2022 \cite{iso27001} and NIST Special Publication (SP) 800-160 \cite{nistSP800}, extending them to a structured mathematical framework.

\textbf{Definition 1:} A system $\mathcal{S}$ is defined as a set of interacting elements organized to achieve one or more stated purposes:
\vspace{-2mm}
\[
\mathcal{S} = \{\mathcal{H}, \mathcal{A}, \mathcal{E}, \mathcal{C}\}.
\]

System $\mathcal{S}$ comprises hosts $\mathcal{H}$, which manage assets $\mathcal{A}$. Each asset $a_k \in \mathcal{A}$ consists of a set of components $\mathcal{C}$, which perform the specific functions required to fulfill the asset's purpose. $\mathcal{E}$ is the set of edges representing relationships and interactions between elements (e.g., communication and dependencies). The hierarchical relationships can be represented as:
\vspace{-2mm}
\[
\mathcal{S} \supset \mathcal{H} \supset \mathcal{A} \supset \mathcal{C}.
\]

\textbf{Definition 2:} A host $h_i \in \mathcal{H}$ is a platform or environment that provides the infrastructure necessary to operate and manage one or more assets. A host can be considered as a set of assets it manages, denoted as:
\vspace{-2mm}
\[
h_i = \{a_{i1}, a_{i2}, \ldots, a_{ip}\} \quad \text{where} \quad a_{ij} \in \mathcal{A}.
\]

Hosts may be physical machines, virtual environments, or any other platform that supports the functionality of the assets.

\textbf{Definition 3:} An asset $a_m \in \mathcal{A}$ is a fundamental element within the system that fulfills specific requirements or functions. An asset may consist of hardware, software, firmware, or a combination of these components:
\vspace{-2mm}
\[
a_m = \{c_{m1}, c_{m2}, \ldots, c_{mn}\}.
\]

The value of an asset is determined by stakeholders based on mission, business, and operational concerns across the entire system lifecycle. Assets are equipped with attributes such as network communication (IP address, MAC address, etc.) and subnet information, which facilitate their interaction with other assets.

\textbf{Definition 4:} A component $c_{mi} \in \mathcal{C}$ is the smallest identifiable element of an asset that performs a specific function. Components can be hardware, software, or firmware, and they are characterized by well-defined inputs and outputs. A component can be mathematically defined as:
\vspace{-2mm}
\[
c_{mi} = \left\{ \text{Input}(c_{mi}), \text{Output}(c_{mi}), \text{Function}(c_{mi}) \right\}.
\]

$\text{Input}(c_{mi})$ and $\text{Output}(c_{mi})$ represent the inputs and outputs of the component. $\text{Function}(c_{mi})$ defines the operation or service provided by the component.

\textbf{Definition 5:} A vulnerability $v_k \in V$ is a weakness or flaw in a component that can be exploited, leading to a security breach. 

Each vulnerability is characterized by various attributes, including CVSS base score $S^{\text{cvss}}_v$, likelihood sub-score $S_{v}^{\text{likelihood}}$, impact sub-score $S_{v}^{\text{impact}}$, the EPSS score $\text{EPSS}(v)$, the existence of exploits $\text{Exploit}(v)$, whether impact changed scope as $\text{ScopeChange}(v)$ and utilized $\text{Ransomware}(v)$:
\vspace{-2mm}
\[
v = \begin{aligned}
& (S^{\text{cvss}}_v, S^{\text{likelihood}}_v, S^{\text{impact}}_v, \text{EPSS}(v), \\
& \text{Exploit}(v), \text{ScopeChange}(v), \text{Ransomware}(v)).
\end{aligned}
\]

\subsection{System Graph Construction}

The system graph $\mathcal{G}$ integrates host, asset, and component interactions. This hierarchical structure is essential for understanding how vulnerabilities at the component level can propagate through assets and hosts, ultimately impacting the entire system. Each level of interaction - host, asset, and component - plays a critical role in determining the system's overall security posture.

Let $\mathcal{G} = (\mathcal{H}, \mathcal{A}, \mathcal{C}, \mathcal{E})$ be the system graph, where:

\begin{itemize}
\item $\mathcal{H} = \{h_1, h_2, \ldots, h_n\}$ represents the set of hosts within the system.

\item $\mathcal{A} = \{a_1, a_2, \ldots, a_m\}$ represents the set of assets within the system.

\item $\mathcal{C} = \{c_1, c_2, \ldots, c_p\}$ represents the components that constitute the assets.

\item $\mathcal{E}$ is the set of edges representing interactions across hosts, assets, and components.
\end{itemize}

Each host $h_i \in \mathcal{H}$ hosts a set of assets $\mathcal{A}(h_i) = \{a_{i1}, a_{i2}, \ldots, a_{ip}\}$. The interactions between hosts are represented by the set of host communication edges:
\vspace{-2mm}
\[
\mathcal{E}_h = \{(h_i, h_j) \mid h_i, h_j \in \mathcal{H}\}.
\]

Each edge \( (h_i, h_j) \) is associated with a weight \( w_h(h_i, h_j) \), indicating the intensity of communication between the hosts.

Assets $a_k \in \mathcal{A}$ hosted on a specific host $h_i$ interact with other assets through network communication and functional dependencies. These interactions are captured by:

\begin{itemize}
\item $\mathcal{E}_c = {(a_i, a_j) \mid a_i, a_j \in \mathcal{A}}$ represents communication edges between assets, characterized by a weight $w_c(a_i, a_j)$, signifying the intensity of network communication between the assets.
\item $\mathcal{E}_f = {(c_i, c_j) \mid c_i \in C(a_k), c_j \in C(a_l), a_k, a_l \in \mathcal{A}}$ represents functional dependencies between components across different assets, with a weight $w_f(c_i, c_j)$.
\end{itemize}

Within each asset \( a_k \), components \( c_i \in C(a_k) \) interact through functional and embedding relationships:

\begin{itemize}
\item $E_f = {(c_i, c_j) \mid c_i, c_j \in C}$ represents functional dependencies between components within the same asset, with a weight $w_f(c_i, c_j)$.
\item $E_e = {(c_i, c_j) \mid c_i, c_j \in C}$ represents embedding relationships between components within the same asset, with a weight $w_e(c_i, c_j)$.
\end{itemize}

\subsection{Component-Level Risk Aggregation}

Component-level risk $\mathbf{R}_{\text{component}, c_i}$ quantifies the risk level of a component by integrating its vulnerability profile and its role within both the asset and the overall system. It is computed as the product of its Component Vulnerability Score (CVS) and Component Centrality, using:
\vspace{-2mm}
\begin{equation}
\begin{aligned}
\label{eq:ComponentRisk}
\mathbf{R}_{\text{component}, c_i} = \mathbf{R}_{\text{component}, c_i}^{\text{CVS}} \times \mathbf{R}_{\text{component}, c_i}^{\text{Centrality}}.
\end{aligned}
\end{equation}

\textbf{Definition 6:} CVS $\mathbf{R}_{\text{component}, c_i}^{\text{CVS}}$ quantifies the security risk of a component $c_i$ based on the CVSS scores of its associated vulnerabilities. 

The Component Vulnerability Score $\mathbf{R}_{\text{component}, c_i}^{\text{CVS}}$ is computed as a weighted sum of the vulnerability scores across different severity levels, adjusted by the number of vulnerabilities, using: 
\vspace{-2mm}
\begin{equation}
\begin{aligned}
\label{eq:CVS}
\mathbf{R}_{\text{component}, c_i}^{\text{CVS}} =  \frac{\sum_{l \in \{C,H,M,L\}} w_l \times \text{Sum}(S_l)}{\sum_{l \in \{C,H,M,L\}} w_l}. 
\end{aligned}
\end{equation}

Here, $w_l$ is the weighting factor for severity level $l$, and $\text{Sum}(S_l)$ is the sum of vulnerability scores within severity scale $l$. Similar weighted aggregation approaches are seen in \cite{ekle2023cyber} and proved in \cite{northern2021vercasm}.

\textbf{Definition 7:} Component centrality $\mathbf{R}_{\text{component}, c_i}^{\text{Centrality}}$ is a measure that reflects the importance of component $c_i$ based on its connectivity and influence within the asset and the system.

$\mathbf{R}_{\text{component}, c_i}^{\text{Centrality}}$ considers a combination of degree centrality, betweenness centrality, and PageRank, and is given by: 
\vspace{-2mm}
\begin{equation}
\begin{aligned}
\label{eq:componentCentrality}
\mathbf{R}_{\text{component}, c_i}^{\text{Centrality}} = \frac{\text{DC}(c_i) + \text{BC}(c_i) + \text{PR}(c_i)}{3}.
\end{aligned}
\end{equation}

The centrality value is normalized to ensure it is in the range [0, 1]. Degree centrality, $\text{DC}(c_i)$, quantifies how frequently a node connects to other nodes in the network, while betweenness centrality, $\text{BC}(c_i)$, captures the proportion of shortest paths between all node pairs that pass through node $i$ \cite{watts1998collective}. PageRank, $\text{PR}(c_i)$ measures the influence of a node based on the number and quality of links to it, emphasizing connections from highly connected components \cite{gleich2015pagerank}.

\subsection{Asset-Level Risk Aggregation}

Asset-level risk $\mathbf{R}_{\text{asset}, a_m}$ is the overall risk score for an asset $a_m$, aggregated from the risks of its components and the vulnerabilities $\mathbf{R}_{\text{vulnerability}, v_k}$ associated with those components, and is given by:
\vspace{-2mm}
\begin{equation}
\begin{aligned}
\label{eq:AssetRisk} 
\mathbf{R}_{\text{asset}, a_m} = \sum_{c_i \in \mathcal{C}(a_m)} \left(\sum_{v_k \in \mathcal{V}(c_i)} \mathbf{R}_{\text{vulnerability}, v_k} \right).
\end{aligned}
\end{equation}

\textbf{Definition 8:} $\mathbf{R}_{\text{vulnerability}, v_k}$ is calculated by considering the direct and indirect risks of vulnerability $v_k$, and defined as:
\vspace{-2mm}
\begin{equation}
\begin{aligned}
\label{eq:VulRisk}
\mathbf{R}_{\text{vulnerability}, v_k} = \mathbf{R}_{\text{direct}, v_k} + \mathbf{R}_{\text{indirect}, v_k}.  
\end{aligned}
\end{equation} 

Direct vulnerability risk $\mathbf{R}_{\text{direct}, v_k}$ quantifies the immediate impact of a vulnerability $v_k$ being exploited on the component $c_i$ to which it is directly associated, scaled by the exploit likelihood $EL_{v_k}$ and the component's centrality within the asset and system, using:
\vspace{-2mm}
\begin{equation}
\begin{aligned}
\label{eq:DirectRisk}
\mathbf{R}_{\text{direct}, v_k} = EL_{v_k} \times S^{\text{impact}}_{v_k} \times \mathbf{R}_{\text{component}, c_i}^{\text{Centrality}}.
\end{aligned}
\end{equation}

\textbf{Definition 9:} Exploit likelihood $EL_{v_k}$ of a vulnerability $v_k$ quantifies the potential for exploitation based on the vulnerability's characteristics and threat intelligence, calculated using:
\vspace{-2mm}
\begin{equation}
\begin{aligned}
\label{eq:ExploitLikelihood}
EL_{v_k} = \alpha \times \left(\frac{S_{v_k}^{\text{likelihood}}}{10}\right) + \beta \times \text{EPSS}(v_k) + \gamma \times \text{Exploit}(v_k).
\end{aligned}
\end{equation}

The weights $\alpha$, $\beta$, and $\gamma$ allow for a customized balance of the influence of these three factors. 

The exploit indicator $\text{Exploit}(v_k)$ is a binary value indicating the existence of an exploit for a vulnerability, defined as:
\vspace{-2mm}
\begin{equation}
\begin{aligned}
\label{eq:ExploitIndicator}
\text{Exploit}(v_k) =
\begin{cases} 
1, & \text{if there is an exploit for vulnerability} \; v_k \\
0, & \text{otherwise}
\end{cases}.
\end{aligned}
\end{equation}

\textbf{Definition 10:} Indirect vulnerability risk, $\mathbf{R}_{\text{indirect}, v_k}$, quantifies the risk that propagates through the network due to interdependencies between components following the exploitation of vulnerability $v_k$.

$\mathbf{R}_{\text{indirect}, v_k}$ is computed as the product of the propagation likelihood $PL_{v_k}$ and the sum of the risks of components impacted by the vulnerability, adjusted by the path weights, as given by:
\vspace{-2mm}
\begin{equation}
\begin{aligned}
\label{eq:PropagatedVulRisk}
\mathbf{R}_{\text{indirect}, v_k} = PL_{v_k} \times \sum_{c_i \in \mathcal{C}_{\text{imp}}} \left( \sum_{(c_j \rightarrow c_i) \in \text{path}} w_{ji} \right) \times S^{\text{impact}}_{v_k}.
\end{aligned}
\end{equation}

This risk is considered only when $PL_{v_k}$ exceeds a predefined threshold $\sigma$ and the initial component $c_i$ has incoming dependencies. The term $\sum_{c_i \in \mathcal{C}_{\text{imp}}}$ represents the summation over all components impacted by vulnerability $v_k$, while $\sum_{(c_j \rightarrow c_i) \in \text{path}}$ indicates the sum of all edge weights along the path from the source component to the affected component. This definitoin is inspired by the works by \cite{pal2020sustainable} \cite{denuit2022conditional} \cite{fumagalli2023semantics}.

\textbf{Definition 11:} Propagation likelihood $PL_{v_k}$ quantifies the probability that the effects of an exploited vulnerability $v_k$ will propagate through the network, calculated by:
\vspace{-2mm}
\begin{equation}
\begin{aligned}
\label{eq:PropagationLikelihood}
PL_{v_k} = \delta \times \text{ScopeChange}(v_k) + \theta \times \text{Ransomware}(v_k).
\end{aligned}
\end{equation}

$\delta$ and $\theta$ are weighting parameters that represent the relative significance of each factor in accordance with the network's security posture and its history of attack patterns.

The scope change indicator, $\text{ScopeChange}(v_k)$, evaluates whether the exploitation of vulnerability $v_k$ results in alterations to the access or interaction scope, given by: 
\vspace{-2mm}
\begin{equation}
\label{eq:SCIndicator}
\text{ScopeChange}(v_k) =
\begin{cases} 
1, & \text{if } v_k \text{ changes scope} \\
0, & \text{otherwise}
\end{cases}.
\end{equation}

Such a change can enable broader access and facilitate the propagation of risk across the network.

The historical ransomware use indicator, $\text{Ransomware}(v_k)$, identifies whether vulnerability $v_k$ has been previously exploited in ransomware attacks, as described by: 
\vspace{-2mm}
\begin{equation}
\label{eq:RVIndicator}
\text{Ransomware}(v_k) =
\begin{cases}
1, & \text{if } v_k \text{ used in ransomware} \\
0, & \text{otherwise}
\end{cases}.
\end{equation}

This indicator suggests the potential for the vulnerability to cause widespread impact, given ransomware's typical strategy of maximizing reach and damage.

\subsection{System-Level Risk Aggregation}

System-level risk $\mathbf{R}_{\text{system}}$ represents the total risk to the system by aggregating both network-based risks and host-based risks, and is given by:
\vspace{-2mm}
\begin{equation}
\begin{aligned}
\label{eq:SystemRisk}
\mathbf{R}{\text{system}} = \mathbf{R}_{\text{network}} + \sum_{h_m \in H} \mathbf{R}_{\text{host}, h_m}.
\end{aligned}
\end{equation}

\textbf{Definition 11:} Host-based risk, denoted as $\mathbf{R}_{\text{host}, h_m}$, represents the aggregate risk associated with all vulnerabilities within the assets hosted on host $h_m$. 

$\mathbf{R}_{\text{host}}$ is calculated by aggregating the risks of all vulnerabilities within the host, weighted by the criticality of the assets in which the vulnerabilities are located: 
\vspace{-2mm}
\begin{equation}
\begin{aligned}
\label{eq:HostRisk}
\mathbf{R}_{\text{host}, h_m} = \sum_{a_m \in \mathcal{A}(h_m)} \mathbf{R}_{\text{asset}, a_m}^{\text{Criticality}} \times \left( \sum_{v_h \in V(a_m)} \mathbf{R}_{\text{vulnerability}, v_h} \right).
\end{aligned}
\end{equation}

The vulnerability risk $\mathbf{R}_{\text{vulnerability}, v_h}$ is determined using Equations (\ref{eq:VulRisk}), (\ref{eq:DirectRisk}), and (\ref{eq:PropagatedVulRisk}), with the indirect risk considering only propagation within the host's internal dependency graph.

\textbf{Definition 12:} Network-based risk, denoted as $\mathbf{R}_{\text{network}}$, represents the risk associated with network-exploitable vulnerabilities propagating across the network between hosts. 

This risk focuses on lateral movement across the network, where vulnerabilities may be exploited to compromise multiple hosts. $\mathbf{R}_{\text{network}}$ is calculated by summing the aggregated risk scores for all shortest paths leading to high-criticality assets: 
\vspace{-2mm}
\begin{equation}
\begin{aligned}
\label{eq:NetworkRisk}
\mathbf{R}_{\text{network}} = \sum_{a_h \in \mathcal{A}_h} \left(\sum_{a \in \text{SAP}(a_s, a_h, \text{path})} \mathbf{R}_{\text{vulnerability}, a}^{\text{network-propagation}}\right).
\end{aligned}
\end{equation}

Here, $\mathcal{A}_h$ refers to the set of high-criticality assets. $\text{SAP}(a_s, a_h, \text{path})$ is the shortest attack path from source $a_s$ to high-criticality asset $a_h$. The risk calculation for each vulnerability $\mathbf{R}_{\text{vulnerability}, a}^{\text{network-propagation}}$ incorporates Equation (\ref{eq:DirectRisk}), while using system-level dependence graph to calculate component centrality (Equation (\ref{eq:componentCentrality})). Vulnerabilities are classified as either host-based or network-based by examining the Attack Vector (AV) field in the CVSS V3 vector. A vulnerability is classified as network-based if its AV metric is labeled as \textit{Network} or \textit{Adjacent}; otherwise, it is considered host-based if the AV metric is labeled as \textit{Local} or \textit{Physical} \cite{cvssAV}.

\textbf{Definition 13:} Asset criticality, denoted as $\mathbf{R}_{\text{asset}, a_m}^{\text{Criticality}}$, is a metric that quantifies the significance of an asset $a_m$ within the system context.

Asset criticality is determined by two factors: the structural importance of the components that comprise the asset (i.e., component centrality) and the business criticality of the asset. The criticality level is calculated using: 
\vspace{-2mm}
\begin{equation}
\label{eq:AssetCriticality}
\begin{aligned}
\mathbf{R}_{\text{asset}, a_m}^{\text{Criticality}} = w_1 \cdot \mathbf{R}_{\text{asset}, a_m}^{\text{Centrality}} + w_2 \cdot \mathbf{R}_{\text{asset}, a_m}^{\text{BusinessCriticality}}.
\end{aligned}
\end{equation}

$w_1$ and $w_2$ are weights that determine the relative importance of centrality and business criticality, respectively. Note that both $\mathbf{R}_{\text{asset}, a_m}^{\text{Centrality}}$ and $\mathbf{R}_{\text{asset}, a_m}^{\text{BusinessCriticality}}$ are normalized to be (0,1]. 

$\mathbf{R}_{\text{asset}, a_m}^{\text{Centrality}}$ aggregates the centrality scores of all components within an asset $a_m$, reflecting the asset's structural importance based on the connectivity and influence of its components. This is calculated using:
\vspace{-2mm}
\begin{equation}
\label{eq:AssetCentrality}
\begin{aligned}
\mathbf{R}_{\text{asset}, a_m}^{\text{Centrality}} = \frac{1}{n_m} \sum_{i=1}^{n_m} \mathbf{R}_{\text{component}, c_i}^{\text{Centrality}}.
\end{aligned}
\end{equation}

Asset business criticality, denoted as $\mathbf{R}_{\text{asset}, a_m}^{\text{BusinessCriticality}}$, is categorized in Table \ref{tab:criticality_levels}. It quantifies the asset's importance to business operations and is built on top of \cite{keskin2021scoring} \cite{kure2022asset} \cite{corallo2020cybersecurity}.

\begin{table}[h]
\centering
\begin{tabular}{|c|l|}
\hline
\textbf{Level} & \textbf{Description} \\
\hline
1 & External/public assets connecting to non-critical data. \\
\hline
2 & Internet-facing assets, such as public web servers. \\
\hline
3 & Internal assets connecting to non-critical data. \\
\hline
4 & Internal assets with business-sensitive applications. \\
\hline
5 & Internal assets with business-critical applications and data. \\
\hline
6 & Internal assets with mission-critical applications essential for immediate operations. \\
\hline
\end{tabular}
\vspace{5pt}
\caption{Asset business criticality levels and descriptions}
\label{tab:criticality_levels}
\end{table}

\textbf{Definition 14:} The shortest attack path (SAP) algorithm identifies the least resistance path susceptible to exploits within a network of assets.

Given a weighted graph $G = (\mathcal{A}, E, w)$, where $\mathcal{A}$ represents the set of vertices (assets), $E$ represents the set of edges (connections between assets), and $w: E \rightarrow \mathbb{R}^+$ assigns a positive weight to each edge (representing the risk or cost of exploitation), the shortest attack path between vertices $a_i$ and $a_j$ is defined as the path $P = \{a_0, a_1, \ldots, a_k\}$ with $a_0 = a_i$ and $a_k = a_j$, such that the sum of the weights of the edges in $P$ is minimized, using:
\vspace{-2mm}
\begin{equation}
\label{eq:ShortestPath}
\begin{aligned}
\text{SAP}(a_i, a_j, \text{path}) = \min \left( \sum_{l=0}^{k-1} w(a_l, a_{l+1}) \right).
\end{aligned}
\end{equation}

\section{Proposed Approach} \label{sec:Approach}

This section presents a comprehensive approach for risk assessment and mitigation in enterprise networks. The methodology begins with extensive data collection, gathering physical, network, software, and threat intelligence data to construct a network communication graph. This foundation enables the retrieval and prioritization of vulnerabilities based on their impact on critical assets. A dependence graph is then constructed to model asset interconnections using functional dependency rules. Asset criticality is evaluated through a multi-dimensional scoring metric that integrates both functional and business factors. Finally, risk scores are calculated by analyzing direct and propagated risks, employing graph-theoretical models to enhance risk prioritization. This holistic approach ensures that patching efforts are optimally allocated to improve overall network security.

\subsection{Network Communication Graph Construction} \label{sec:dataCollection}

The network communication graph and dependence graph form the foundation of our system-wide risk assessment and vulnerability analysis approach. We employ a top-down methodology for data collection, starting with a comprehensive inventory of network devices, including firewalls, switches, and routers. This is followed by the analysis of network and firewall logs to identify internal communication flows. These graphs enable accurate modeling of potential attack paths and risk propagation.

Our asset management and system configuration data collection architecture integrates tools such as \textit{Ansible}, \textit{Node.js}, \textit{React}, and \textit{VMWare}. The process begins with IT asset registration through a scalable web-based system. \textit{Ansible} automates asset information extraction via secure WinRM or SSH connections, employing an agentless approach to minimize intrusion while gathering detailed configuration data. This method efficiently collects information on operating systems, installed software, and running services. Nevertheless, as this paper focuses on risk aggregation and vulnerability ranking, a detailed discussion of the asset management process is beyond the scope.

The extracted data is then formatted as JSON files and stored in a MongoDB database for further analysis, with the keys listed in Table \ref{tab:GlobalFeatures}. 

\begin{table}[!ht]
\centering
\caption{Extracted Key Features from Asset Management}
\label{tab:GlobalFeatures}
\begin{tabular}{|l|l|}
\hline
\textbf{Feature}            & \textbf{Description}                                               \\ \hline
\multicolumn{2}{|c|}{\textbf{Asset-Level Features}}                                              \\ \hline
Asset Type                  & Category of the asset (e.g., Workstation)                  \\ \hline
Operating System            & OS running on the asset (e.g., Windows)                     \\ \hline
IP Address                  & Asset’s IPv4 or IPv6 address                                       \\ \hline
MAC Address                 & Media access control address                                       \\ \hline
Patch Level                 & Installed patches (e.g., KB numbers)                               \\ \hline
Running Services            & Active services on the asset                                       \\ \hline
Configuration Details       & Paths and settings                                                 \\ \hline
\multicolumn{2}{|c|}{\textbf{Component-Level Features}}                                          \\ \hline
Vendor                      & Manufacturer of the component                                      \\ \hline
Product Name                & Name of the installed component                                    \\ \hline
Version                     & Version identifier of the component                                \\ \hline
\end{tabular}
\end{table}

Next, we introduce how to integrate system configuration with software and cyber threat intelligence data.

\subsection{Vulnerability Information Retrieval} \label{sec:dataCollection} 

We leverage the detailed asset and component information, specifically the vendor, product name, version, and the relationships between components (such as which component is running on or with another component or operating system), to generate Common Platform Enumeration (CPE) 2.3 metadata. For instance, a server running Apache HTTP Server 2.4.41 on Ubuntu Linux 20.04 would be represented by the CPE 2.3 strings \textit{cpe:2.3:o:canonical:ubuntu\_linux:20.04} and \textit{cpe:2.3:a:apache:http\_server:2.4.41}. This standardized metadata facilitates accurate vulnerability matching and retrieval.

Using such generated CPE 2.3 metadata, we can query the NVD through its API \cite{nvd-api} to retrieve a comprehensive list of applicable vulnerabilities. Additionally, we utilize \textit{Shodan} to identify vulnerabilities associated with public-facing assets by leveraging IP addresses. According to the Mandiant M-Trends 2024 report, exploits are a prevalent initial infection vector, responsible for 38\% of intrusions \cite{Mandiant24}. To address this, we also cross-check databases like \textit{ExploitDB} and \textit{CVEDetails} to gather critical information on available exploits and ransomware associated with vulnerabilities. This approach supports a dynamic analysis of the evolving threat landscape, ensuring that we collect relevant, up-to-date vulnerability data tailored to our system’s unique components and configurations.

The following subsections detail the components of our proposed multi-metric scoring system and its application in vulnerability prioritization.

\subsection{Dependence Graph Construction} \label{sec:Dependence}

This step involves processing extracted data (introduced in Section \ref{sec:dataCollection}) to construct a graph that represents the interconnections and dependencies within an enterprise network. This graph forms the foundational structure for further analysis, allowing us to model and simulate potential vulnerability exploitation paths and assess the impact of various security vulnerabilities. We utilize software configuration data to identify dependencies within assets, and use the network communication data to identify dependencies between assets.

More specifically, we employ an integrated approach combining functional dependency (FD) rules to generate a dependence graph, which is extended on top of our previous works in \cite{jiang2023model} and \cite{jiang2024leveraging}. The notation $FD_{(i,j)}$ suggests that component $i$ depends on component $j$. FD rules provide a qualitative basis by categorizing connections according to their functional roles within the system architecture. FD rules are further categorized into embedding, interaction, data, service, security and network rules. Network FD rules are applicable to asset dependencies, while the other rules normally exist between components. Interaction rules often involve active data exchanges required for operational control or processing and are particularly relevant for ICS, whereas data rules are about passive data reception or monitoring.

\begin{enumerate}
\item \textit{FD Embedding Rule (ER):}
\begin{itemize}
\item \textit{ER-1:} $FD_{(i,j)} \leftarrow C_i \subseteq C_j$, for $C_i$  embedded in $C_j$. 
\item \textit{ER-2:} $FD_{(j,i)} \leftarrow C_j \subseteq C_i$, for $C_j$  contained in $C_i$. 
\end{itemize}

\item \textit{FD Interaction Rule (IR):}
\begin{itemize}
\item \textit{IR-1:} $FD_{(i,j)} \leftarrow C_i \xleftarrow[]{\text{process}} C_j$, for $C_i$ receiving process data from $C_j$. 
\item \textit{IR-2:} $FD_{(j,i)} \leftarrow C_j \xleftarrow[]{\text{control}} C_i$, for $C_j$ receiving control data from $C_i$. 
\end{itemize}

\item \textit{FD Data Rule (DR):}
\begin{itemize}
\item \textit{DR-1:} $FD_{(i,j)} \leftarrow C_i \xleftarrow[]{\text{stream}} C_j$, for $C_i$ as a data stream recipient from $C_j$. 
\item \textit{DR-2:} $FD_{(i,j)} \leftarrow C_i \xleftarrow[]{\text{listen}} C_j$, for $C_i$ listening to the data stream from $C_j$. 
\end{itemize}

\item \textit{FD Service Rule (SR):}
\begin{itemize}
\item \textit{SR-1:} $FD_{(i,j)} \leftarrow C_i \xleftarrow[]{\text{service}} C_j$, for $C_j$  providing specific services or functionalities to $C_i$. 
\end{itemize}

\item \textit{FD Security Rule (SCR):}
\begin{itemize}
\item \textit{SCR-1:} $FD_{(i,j)} \leftarrow C_i \xleftarrow[]{\text{security}} C_j$, for $C_i$ receiving security controls from $C_j$. 
\end{itemize}

\item \textit{FD Network Rule (NR):}
\begin{itemize}
\item \textit{NR-1:} $FD_{(i,j)} \leftarrow A_i \hookrightarrow[\text{network}] A_j$, for $A_i$ relies on $A_j$ for network connectivity or routing purposes.

\end{itemize}

\end{enumerate}

$\subseteq$ denotes an embedding relationship, where one component is embedded within another. $\xleftarrow[]{\text{process}}$ and $\xleftarrow[]{\text{control}}$ represent data flow and control flow dependency, respectively, where one component receives process or control data from another. Similarly, $\xleftarrow[]{\text{stream}}$ indicates a data stream relationship, where one component is the recipient of a data stream from another. $\xleftarrow[]{\text{listen}}$ indicates a listening relationship, where one component listens to a data stream from another. $\xleftarrow[]{\text{service}}$ and $\xleftarrow[]{\text{security}}$ represent service or security control dependency, respectively. $\hookrightarrow[\text{network}]$ symbolizes network connectivity, where one component is connected to the network via another.

We encode these FD rules to generate dependence connections in the graph, namely $E_f = {(c_i, c_j) \mid c_i, c_j \in C}$ and $E_e = {(c_i, c_j) \mid c_i, c_j \in C}$ within the same assets, as well as \\
$\mathcal{E}_f = {(c_i, c_j) \mid c_i \in C(a_k), c_j \in C(a_l), a_k, a_l \in \mathcal{A}}$ between components across different assets. Practical weights for edges in the dependence graph are assigned based on the criticality of each type of dependency. Embedding and network dependencies are given higher weights (2) due to their foundational role in system functionality, while interaction, data, service, and security dependencies are assigned moderate weights (1), reflecting their importance in operational contexts.

The constructured dependency graph is used to calculate asset centrality, which is detailed in the next sub-section.

\subsection{Asset Criticality Calculation} \label{sec:Criticality}

Asset criticality is determined based on data sensitivity, operational importance of supported applications, and network communication role. Data sensitivity reflects the potential harm from compromising sensitive information, while application criticality indicates the asset's importance to organizational operations and strategic objectives. The asset's network role considers its significance in data transmission and processing. A multi-dimensional scoring metric, integrating both technical and business factors, is employed for quantitative criticality assessment. This approach combines normalized centrality scores with business criticality levels using Equation (\ref{eq:AssetCriticality}). The centrality scores reflect the asset's network importance derived from connectivity and influence. Table \ref{tab:criticality_levels} outlines the business criticality levels, which is built on top of \cite{kure2022asset} \cite{corallo2020cybersecurity}. This comprehensive evaluation enhances the precision of criticality assessment, providing a more accurate basis for resource allocation in vulnerability management.

\subsection{Risk Score Calculation} \label{sec:CalRisk}

Our approach innovatively employs dual graph-theoretical models to enhance risk analysis: a dependence graph analyzes impact by examining component relational dependencies, while a network communication graph identifies shortest attack paths, optimizing critical vulnerability and exposure point detection.

Asset risk scores are computed by synthesizing direct and indirect risks from vulnerabilities in each system component. This process evaluates exploitation likelihood, consequential system impacts, and risk diffusion across interconnected components, as detailed in Algorithm \ref{alg:CalculateAssetRisk}.

\begin{algorithm}[h]
\SetAlgoLined
\KwIn{Processed Data $D$, Parameters $\alpha$, $\beta$, $\sigma$, $\delta$, $\theta$, $\gamma$}
\KwOut{Asset risk $\mathbf{R}_{\text{asset}, a_m}$}
\SetKwFunction{FCalculateComponentRisk}{CalCompRisk}
\SetKwFunction{FPropagateRisks}{IndirectRisk}
\SetKwFunction{FCalculateAssetRisk}{AssetRisk}
\SetKwFunction{FCalculateExploit}{Exploit}
\SetKwFunction{FCalculatePropagation}{Propagation}
\SetKwProg{Fn}{Function}{:}{\KwRet}
\Fn{\FCalculateAssetRisk{$D, \alpha, \beta, \sigma, \delta, \theta, \gamma$}}{
    $C \gets D[\text{Components}]$
    
    $V \gets D[\text{Vulnerabilities}]$
    
    $\mathbf{R}_{\text{asset}, a_m} \gets 0$

    \ForEach{$v_k \in V$}{
        $c_s \gets v_k.\text{component}$
        
        $EL_{v_k} \gets \FCalculateExploit(S^{\text{likelihood}}_{v_k}, \text{EPSS}({v_k}), \text{Exploit}({v_k}), \alpha, \beta, \gamma)$
        
        $\mathbf{R}_{\text{direct}, v_k} \gets EL_{v_k} \times S^{\text{impact}}_{v_k} \times \mathbf{R}_{\text{component}, c_s}^{\text{Centrality}}$
        
        $PL_{v_k} \gets \FCalculatePropagation(v_k, \delta, \theta)$
        
        \If{$PL_{v_k} \geq \sigma$}{
            $\mathbf{R}_{\text{vulnerability}, v_k} \gets \FPropagateRisks(c_s, S^{\text{impact}}_{v_k}, PL_{v_k}, C)$
        } 
        \Else {
            $\mathbf{R}_{\text{vulnerability}, v_k} \gets \mathbf{R}_{\text{direct}, v_k}$
        }
        $\mathbf{R}_{\text{asset}, a_m} \gets \mathbf{R}_{\text{asset}, a_m} + \mathbf{R}_{\text{vulnerability}, v_k}$
    }
    
    \KwRet $\mathbf{R}_{\text{asset}, a_m}$
}

\Fn{\FPropagateRisks{$c_s, \text{impact}, PL_{v_k}, C$}}{
    Queue $Q \gets [(c_s, 0)]$
    $visited \gets \{\}$
    $\mathbf{R}_{\text{total}} \gets \mathbf{R}_{\text{direct}, v_k}$
    
    \While{$Q$ is not empty}{
        $(c_i, \text{weight}) \gets Q.\text{dequeue}()$
        
        \If{$c_i \notin visited$}{
            $visited.\text{add}(c_i)$
            
            \ForEach{$c_j \in C$ \textbf{and} $c_j \notin visited$}{
                \If{$w_{ji} > 0$}{
                    $new\_weight \gets weight + w_{ji}$
                    
                    $new\_impact \gets \text{impact} \times new\_weight$
                    
                    $\mathbf{R}_{\text{total}} \gets \mathbf{R}_{\text{total}} + new\_impact$
                    $Q.\text{enqueue}(c_j, new\_weight)$
                }
            }
        }
    }
    $\mathbf{R}_{\text{total}} \gets \mathbf{R}_{\text{total}} \times PL_{v_k}$
    \KwRet $\mathbf{R}_{\text{total}}$
}

\caption{Calculate Asset Risk}
\label{alg:CalculateAssetRisk}
\end{algorithm}

The methodology begins by calculating vulnerability exploit likelihood using a weighted formula combining normalized CVSS likelihood, EPSS score, and known exploit presence. Weights $\alpha$, $\beta$ and $\gamma$ allow context-specific tuning of the asset. Direct risk is computed as the product of exploit likelihood and vulnerability impact score. For vulnerabilities where the scope of impact has changed (\textquotedblleft \textit{ScopeChanged}\textquotedblright = \textit{True}), the methodology includes a risk propagation step. This involves examining the asset's dependency graph to simulate how vulnerabilities can indirectly affect other components. The adjacency matrix quantifies these dependencies, enabling calculation of indirect risks that are aggregated with direct risks to derive the total asset risk score.

The system-level risk, $\mathbf{R}_{\text{system}}$, provides assessment of network security by aggregating host-based and network-based risks. In our proposed approach, host-based risk calculations focus on direct and internal impacts of vulnerabilities within individual hosts, considering both host-based and network-based vulnerabilities but limiting scope to internal propagation. Network-based risk calculations exclusively address cross-host propagation of network-exploitable vulnerabilities. This separation prevents overlapping calculations, ensuring accurate risk attribution and avoiding double-counting.

Algorithm \ref{alg:CalculateSystemRisk} details the calculation process, employing graph-based methodologies to assess and prioritize risks. Critical assets are identified using a threshold parameter $\gamma$, filtering assets based on criticality levels. The Shortest Attack Path (SAP) methodology, utilizing Dijkstra's algorithm \cite{ortalo1999experimenting}, computes the shortest paths from all assets to these critical assets, identifying routes most susceptible to exploitation while focusing on routes that could facilitate lateral movement within the network. For each path, the algorithm calculates cumulative network-based risk by summing risks of network-exploitable vulnerabilities along the path. The total system-level risk is determined by aggregating risks from both host-based and network-based calculations, providing a global perspective on network security vulnerabilities.

\begin{algorithm}[h]
\SetAlgoLined
\KwIn{Processed Data $D$, Parameters $\alpha$, $\beta$, $\sigma$, $\gamma$, $\delta$, $\theta$}
\KwOut{System-level risk $\mathbf{R}_{\text{system}}$}
\SetKwFunction{FIdentifyCriticalAssets}{CriticalAssets}
\SetKwFunction{FShortestPaths}{ShortestPaths}
\SetKwFunction{FSystemRisk}{SystemRisk}
\SetKwFunction{FCalculateAssetRisk}{CalculateAssetRisk}
\SetKwFunction{FNetworkRisk}{NetworkRisk}
\SetKwProg{Fn}{Function}{:}{\KwRet}
\Fn{\FSystemRisk{$D, \alpha, \beta, \sigma, \gamma, \delta, \theta $}}{
    $\mathcal{A} \gets D[\text{Assets}]$
    
    Initialize $\mathbf{R}_{\text{system}} \gets 0$, $\mathbf{R}_{\text{host}} \gets 0$, $\mathbf{R}_{\text{network}} \gets 0$

    Initialize critical assets $\mathcal{A}_h \gets []$
    
\ForEach{$a_i \in \mathcal{A}$}{
    $\mathbf{R}_{\text{asset}, a_i} \gets \FCalculateAssetRisk(a_i, \alpha, \beta)$
    
    $\mathbf{R}_{\text{host}} \gets \mathbf{R}_{\text{host}} + \mathbf{R}_{\text{asset}, a_i}$
    
        \If{$a_i[\text{criticality\_level}] > \gamma$}{
            $\mathcal{A}_h$.append($a_i$)
        }
    }
$\text{SAP} \gets \FShortestPaths(D, \mathcal{A}_h)$

$\mathbf{R}_{\text{network}} \gets \FNetworkRisk(\text{SAP}, D, \alpha, \beta, \sigma)$

$\mathbf{R}_{\text{system}} \gets \mathbf{R}_{\text{network}} + \mathbf{R}_{\text{host}}$

\KwRet $\mathbf{R}_{\text{system}}$

}

\SetKwProg{Fn}{Function}{:}{\KwRet}
\Fn{\FShortestPaths{$D, \mathcal{A}_h$}}{
    Initialize $paths \gets []$
    
    \ForEach{$a_h$ in $\mathcal{A}_h$}{
        \ForEach{$a_j$ in $\mathcal{A}$}{
            \If{$a_j \neq a_h$}{
                $\mathcal{A}_\text{path} \gets \text{Shortest path from } a_j \text{ to } a_h$
                $paths.append((a_j, a_h, \mathcal{A}_\text{path}))$
            }
        }
    }
    \KwRet $paths$
}

\SetKwProg{Fn}{Function}{:}{\KwRet}
\Fn{\FNetworkRisk{$SAP, D, \alpha, \beta, \sigma$}}{
    Initialize $\mathbf{R}_{\text{network}} \gets 0$
    
        \ForEach{$(a_s, a_h, \mathcal{A}_\text{path})$ in $\text{SAP}$}{
        
        \ForEach{$a_i \in \mathcal{A}_\text{path}$}{
           
            \ForEach{$v_k \in a_i[\text{vulnerabilities}]$}{
                \If{$v_k^{\text{AV}}$ == \textquotedblleft \textit{N}\textquotedblright OR $v_k^{\text{AV}}$ == \textquotedblleft \textit{A}\textquotedblright}{
                $c_s \gets v_k.\text{component}$

                $EL_{v_k} \gets \FCalculateExploit(v_k, \alpha, \beta, \gamma)$
                
        
        $\mathbf{R}_{\text{vulnerability}, v_n}^{\text{network}} \gets EL_{v_k} \times S^{\text{impact}}_{v_k} \times \mathbf{R}_{\text{component}, c_s}^{\text{Centrality}}$
                    
                    $\mathbf{R}_{\text{network}} \gets \mathbf{R}_{\text{network}} + \mathbf{R}_{\text{vulnerability}, v_n}^{\text{network}}$
                }
            }
        }
    }
    \KwRet $\mathbf{R}_{\text{network}}$
}

\caption{Calculate System-Level Risk}
\label{alg:CalculateSystemRisk}
\end{algorithm}

The computational efficiency and scalability of the risk assessment algorithms are crucial for real-time and large-scale network applications. Here we provide an analysis of the time complexity: 

\begin{itemize}
    \item Asset risk calculation: The complexity is approximately $O(V \cdot (C + E))$, where $V$ is the number of vulnerabilities per asset, $C$ is the number of components per asset, and $E$ is the number of edges in the component dependence graph. This complexity arises from the need to perform Breadth-First Search (BFS) for each vulnerability, particularly in scenarios where risk propagation needs to be evaluated across the asset's graph structure.
    \item System-level risk aggregation: The complexity is primarily determined by the shortest path computation and the subsequent path-based risk aggregation. In sparse networks, where the number of edges $E$ is significantly lower than $A^2$, the shortest path computation complexity is $O(A^2 \log A + E)$. However, in dense networks, where $E \approx A^2$, this complexity escalates to $O(A^3 \log A)$. In addition, the path aggregation process, which calculates the total risk contribution along identified paths, has a time complexity of \(O(A^3)\). The overall time complexity of the system-level risk calculation is therefore $O(A^3 \log A)$ when the network is dense, or $O(A^2 \log A + E)$ when the network is sparse.

\end{itemize}

These complexities indicate that the algorithm scales linearly with the number of vulnerabilities and components, while system-level risk aggregation leverages efficient path computation techniques. This ensures that the approach remains feasible for large-scale network assessments.

\subsection{Patch Prioritization} \label{sec:patchRank}

The patch prioritization module aims to identify and rank patches based on their effectiveness in reducing the overall risk score of an asset. We first calculate the risk reduction achieved by removing the vulnerability, and then we rank patches based on the magnitude of risk reduction, using:
\vspace{-2mm}
\begin{equation}
\begin{aligned}
\label{eq:PatchPrioritize}
\text{Rank}_{ij} = \text{sort}(R_{\text{initial}} - R_{\text{patched}}^{(ij)}).
\end{aligned}
\end{equation}

Component-level patch prioritization ranks vulnerabilities based on three key factors that contribute to vulnerability risk: the vulnerability impact $S^{\text{impact}}_{v_k}$, exploit likelihood $EL_{v_k}$ and propagation likelihood $PL_{v_k}$. At the asset level, patch prioritization involves a comprehensive evaluation of vulnerabilities to determine the potential reduction in risk that can be achieved through remediation. This step is essential for understanding how individual patches contribute to improving the security posture of specific assets.

The patch prioritization process extends to a system-wide scope by aggregating the risk reduction associated with vulnerability patching across the entire network. Vulnerabilities are ranked based on their overall impact on system-level risk, ensuring that those posing the greatest threat to system security are addressed first. This structured approach enables efficient resource allocation and effective risk mitigation, ultimately enhancing the resilience and security of the infrastructure.

\section{Case Study} \label{sec:CaseStudy}

\subsection{A Scenario of Enterprise Network} \label{sec:Scenario}

\begin{figure}[h]
\centering
\includegraphics[width=0.6\textwidth]{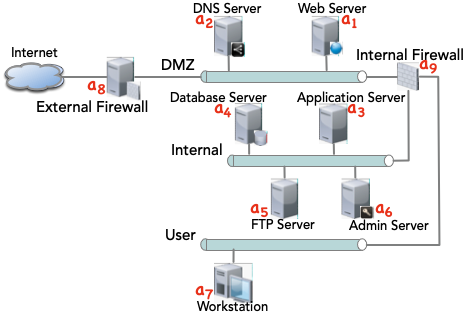}
\caption{Example scenario of an enterprise network configuration.}
\label{fig:example_scenario}
\end{figure}

Consider an enterprise network consisting of multiple interconnected subnets, each protected by internal and external firewalls. This network is designed to support critical business functions and includes various types of servers and workstations, as visualized in Fig. \ref{fig:example_scenario}. The network topology and configuration have been meticulously crafted based on comprehensive analyses of real-world enterprise network environments, with particular attention given to the authoritative reports published by the Cybersecurity and Infrastructure Security Agency (CISA) \cite{cisa-repo}.

\begin{itemize}
    \item External Firewall ($a_8$) segregates the Internet from the enterprise's Demilitarized Zone (DMZ), controlling inbound and outbound traffic to mitigate external threats.
    \item DMZ Subnet hosts the Web Server ($a_1$) and DNS Server ($a_2$), providing services to external requests while isolating direct access to the internal network. 
    \item Internal Firewall ($a_9$) provides additional protection for the internal critical assets, guarding against potential threats that may propagate from the DMZ. 
    \item Internal Subnet hosts assets such as the Application Server ($a_3$), Database Server ($a_4$), FTP Server ($a_5$), and Administrative Server ($a_6$). This subnet is designed to secure crucial data and processes.
    \item User Subnet includes user workstations ($a_7$), used by employees for daily operations, accessing resources within the DMZ and internal subnets.
\end{itemize}

\begin{figure}[h]
\centering
\includegraphics[width=0.6\textwidth]{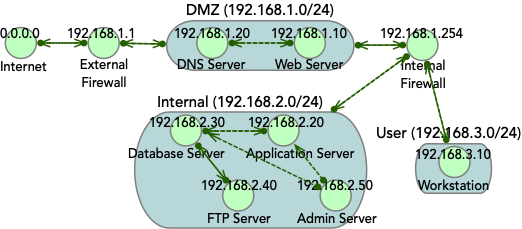}
\caption{Example of network communication graph.}
\footnote{Note that the IP addresses are only for illustration.}
\label{fig:example_communication}
\end{figure}

We begin by constructing a network communication graph, as illustrated in Fig. \ref{fig:example_communication}. The DMZ zone contains servers (DNS and Web) configured to handle direct Internet requests. The Internal Firewall controls data flow between DMZ and internal assets, ensuring only authorized communications traverse these zones. For instance, \textquotedblleft Web Server $\rightarrow$ Internal Firewall $\rightarrow$ Application Server\textquotedblright \space represents the connection for processing web requests. The Application Server communicates with various servers to facilitate controlled data flow for business operations. The Workstation connects to the User Network via the Internal Firewall. The Workstation is connected directly to the Internal Firewall, which serves as a gateway to the User Network. Note that the IP addresses in Fig. \ref{fig:example_communication} are set for demonstration purpose only.

Vulnerabilities applicable to the system are then retrieved and integrated with features such as CVSS vectors and exploit counts. In the enterprise network (shown in Fig. \ref{fig:example_scenario}), several vulnerabilities have been identified. The Web Server ($a_1$), utilizing Apache HTTP Server and Tomcat, is associated with several vulnerabilities. These include CVE-2022-0492 and CVE-2020-25719 (privilege escalation in Ubuntu 20.04); CVE-2021-3156 (sudo utility), CVE-2021-41773 (path traversal in Apache HTTP Server); CVE-2020-1938, CVE-2024-21733, and CVE-2023-41080 (affecting Apache Tomcat).

The Application Server ($a_3$), a critical component in the internal subnet, operates on Debian Linux 12.0 and hosts Oracle WebLogic Server. It is vulnerable to CVE-2024-6387 and CVE-2023-47272 (OS-level); CVE-2023-21839, CVE-2022-22965, and CVE-2024-20986 (WebLogic Server); CVE-2021-2163 and CVE-2021-2161 (Java SE Runtime Environment), and CVE-2021-41773 (embedded Apache HTTP Server). This complex vulnerability landscape demonstrates the challenges in prioritization, considering the interconnected nature of components and the diverse attack vectors they present.

The complete information of the vulnerabilities across the enterprise network is detailed in Table \ref{tab:completeRank_I} in the Appendix.

\subsection{Construct Dependence Graph and Calculate Criticality}

\begin{figure}[h]
\centering
\includegraphics[width=0.6\textwidth]{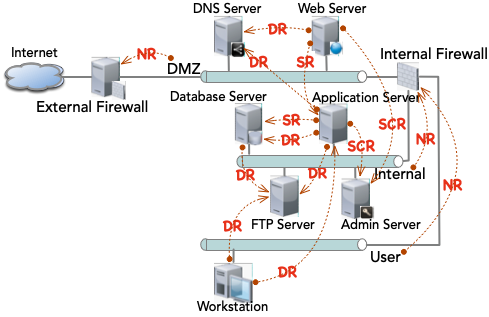}
\caption{Example of dependency analysis.}
\label{fig:dependence}
\end{figure}

We follow the approaches introduced in Sections \ref{sec:Dependence} and \ref{sec:Criticality} to construct system-wide dependence graph as illustrated in Fig. \ref{fig:dependence}, which is utilized to calculate component centrality, asset centrality and asset criticality. 

Using the Web Server ($a_1$) as a case study, we explore its dependencies beyond embedding relationships. The Apache HTTP Server ($c_2$) operates on Ubuntu 20.04 ($c_1$) and forwards incoming requests to Apache Tomcat ($c_3$) for processing. We define a service dependency $SR_{\text{(HTTP}(a_1), \text{Tomcat}(a_1))}$, reflecting Tomcat's role as an independent backend processor operating in conjunction with the HTTP Server for complete web application service. This leads to the following adjacency matrix:
\vspace{-2mm}
\[
\text{Adjacency Matrix of } a_1 =
\begin{bmatrix}
0 & 0 & 0  \\
2 & 0 & 1  \\
2 & 0 & 0  \\
\end{bmatrix}.
\]

The Web Server functions as a gateway, forwarding requests to the Oracle WebLogic Server on the Application Server ($a_3$) for business logic processing, defined as a cross-asset interaction dependency $SR_{\text{(HTTP}(a_1), \text{WebLogic}(a_3))}$. It also relies on the DNS Server ($a_2$) for domain name resolution, represented as a data dependency $DR_{\text{(HTTP}(a_1), \text{DNS}(a_2))}$. Additionally, the Web Server ($a_1$) depends on the Administration Server ($a_6$) for security settings and traffic monitoring, denoted by the security dependency $SCR_{\text{(HTTP}(a_1), \text{ConfigMgr}(a_6))}$.

Expanding to the Application Server ($a_3$), it operates on Debian Linux ($c_1$), with Oracle WebLogic Server ($c_2$) relying on Java SE Runtime ($c_3$) for Java application execution. The Apache HTTP Server ($c_4$) depends on WebLogic Server for dynamic content handling through proxy setups, as summarized in the following adjacency matrix:
\vspace{-2mm}
\[
\text{Adjacency Matrix of } a_3 =
\begin{bmatrix}
0 & 0 & 0 & 0 \\
2 & 0 & 1 & 0 \\
2 & 0 & 0 & 0 \\
2 & 1 & 0 & 0 \\
\end{bmatrix}.
\]

The Application Server's dependencies extend beyond $SR_{\text{(HTTP}(a_1), \text{WebLogic}(a_3))}$. It relies on the Database Server ($a_4$) for data retrieval and storage, forming service and data dependencies \\
$SR_{\text{(WebLogic}(a_3), \text{SQL}(a_4))}$ and $DR_{\text{(WebLogic}(a_3), \text{SQL}(a_4))}$. File transfers with the FTP Server ($a_5$) are denoted as $DR_{\text{(WebLogic}(a_3), \text{FileZilla}(a_5))}$. DNS Server ($a_2$) enhances the Application Server's performance through domain resolution, represented by $DR_{\text{(WebLogic}(a_3), \text{DNS}(a_2))}$. Security measures are maintained via the Administration Server ($a_6$), indicated by $SCR_{\text{(WebLogic}(a_3), \text{ConfigMgr}(a_6))}$.

The network architecture is protected by two firewalls. The External Firewall ($a_8$) demarcates the DMZ, regulating internet-internal network traffic, denoted as $NR_{\text{(DMZ, Firewall}(a_8))}$. The Internal Firewall ($a_9$) safeguards the core network and user subnet, represented by $NR_{\text{(Internal, Firewall}(a_9))}$ and $NR_{\text{(User, Firewall}(a_9))}$.

Asset criticality is calculated using the constructed dependence graph. We integrate normalized centrality and business criticality scores through a weighted sum (Equation (\ref{eq:AssetCriticality})). The final score is converted to an integer scale (0-10) using  $\left\lfloor \mathbf{R}_{\text{asset}, {a_m}}^{\text{Criticality}} \times 10 \right\rfloor$. For example, the Web Server, an Internet-facing asset with business criticality Level 2 ($\mathbf{R}_{\text{asset}, {webServer}}^{\text{BusinessCriticality}} = 2$), has a normalized centrality score of 0.8989 and normalized business criticality of 0.3. Assuming weights $(w_1 = 0.6, w_2 = 0.4)$, the final criticality score is $0.6 \times 0.3572 + 0.4 \times 0.3 = 0.3343$, or 3 in integer format.

\subsection{Risk Calculation and Vulnerability Ranking}

We analyze the Web Server as a case study to demonstrate the asset risk score calculation process. We use weights $\alpha = 0.3$, $\beta = 0.4$, and $\gamma = 0.3$ for exploit likelihood calculation, and $\delta = 0.5$ and $\theta = 0.5$ for propagation likelihood calculation.

\begin{itemize}
    \item Ubuntu component contains three vulnerabilities: \textit{CVE-2022-0492}, \textit{CVE-2020-25719}, and \textit{CVE-2021-3156}, with CVSS scores of 7.8, 7.2, and 7.8, respectively. Calculated exploit likelihood values are 1.8 for \textit{CVE-2022-0492} and \textit{CVE-2021-3156}, and 1.2 for \textit{CVE-2020-25719}, resulting in direct risks of 2.3132, 0.2153, and 4.3573, respectively. These vulnerabilities do not contribute to indirect risk propagation due to their \textquotedblleft \textit{ScopeChanged}\textquotedblright \space and \textquotedblleft \textit{Ransomeware}\textquotedblright \space attributes being false.
    
    \item  Apache HTTP Server component has one vulnerability, \textit{CVE-2021-41773}, with a direct risk of 2.3980. While its \textquotedblleft \textit{Ransomeware}\textquotedblright \space attribute is true, the absence of incoming dependencies limits its indirect impact.
    
    \item Apache Tomcat component contains multiple vulnerabilities. Among them, vulnerability \textit{CVE-2020-1938} has a high CVSS score (9.8), resulting in a significant direct risk of 4.1423. \textit{CVE-2023-41080} and \textit{CVE-2019-0221} have the \textquotedblleft \textit{ScopeChanged}\textquotedblright \space attribute set to true, indicating indirect risk propagation with a score of 1.35 each, leading to risk scores of 1.5498 and 2.2633, respectively.
\end{itemize}

The Web Server's overall asset risk aggregates to 16.6895, which is not a direct sum of component risks (Ubuntu: 6.8400, Apache HTTP Server: 2.2500, Apache Tomcat: 9.6700).

Following the same process, we compute the risk scores for other assets as: DNS Server (7.7409), Application Server (20.8915), Database Server (19.2248), FTP Server (7.8868), Administrative Server (23.3039), User Workstation (20.9366), External Firewall (2.2669), and Internal Firewall (0.5915).

With a criticality threshold $\gamma = 4$, the critical asset set $A_h$ = \{\text{Application Server}, \text{Database Server}\}. The SAP algorithms identify two shortest paths to these critical assets from the Internet. For example, the path to the Application Server is: Internet $\rightarrow$ External Firewall $\rightarrow$ Web Server $\rightarrow$ Internal Firewall $\rightarrow$ Internal Network Gateway $\rightarrow$ Application Server. The path to the Database Server extends this by Application Server $\rightarrow$ Database Server. Summing vulnerability risks along these paths yields a network-based risk score of 20.2119. Combined with the total host-based risk of 51.6504, the final system-level risk score is 71.8623.

Vulnerability ranking for patch prioritization is based on risk reduction effectiveness, as discussed earlier in Section \ref{sec:patchRank}. Using the Web Server as a case study, the top three vulnerabilities identified for patching are listed below. Although vulnerability \textit{CVE-2021-3156} has a lower CVSS score than \textit{CVE-2020-1938}, its higher centrality attribute from $c_1$ results in a significant risk reduction of 4.357 when patched. To enhance the explainability of our patch prioritization ranking and support cybersecurity decision-making, we provide detailed attributes for each vulnerability.

\begin{itemize}
    \item \textbf{CVE-2021-3156} (Exists in $c_1$)
    \begin{itemize}
        \item CVSS: 7.8 (Likelihood: 1.8; Impact: 5.9)
        \item EPSS: 0.96133
        \item Scope changed: False
        \item Utilized Ransomware: 0
        \item Existing exploit: 1
    \end{itemize}
\end{itemize}

\begin{itemize}
    \item \textbf{CVE-2020-1938} (Exists in $c_3$)
    \begin{itemize}
        \item CVSS: 9.8 (Likelihood: 3.9; Impact: 5.9)
        \item EPSS: 0.97284
        \item Scope changed: False
        \item Utilized Ransomware: 0
        \item Existing exploit: 1
    \end{itemize}
\end{itemize}

\begin{itemize}
    \item \textbf{CVE-2021-41773} (Exists in $c_2$)
    \begin{itemize}
        \item CVSS: 7.5 (Likelihood: 3.9; Impact: 3.6)
        \item EPSS: 0.97447
        \item Scope changed: False
        \item Utilized Ransomware: 1
        \item Existing exploit: 1
    \end{itemize}
\end{itemize}

The patch ranking for vulnerabilities across the entire enterprise network is detailed in Table \ref{tab:completeRank_I}. The top three vulnerabilities identified are \textit{CVE-2023-36884} in the Windows 2019 Server on the Database Server, \textit{CVE-2023-47272} in the Debian Linux OS on the Application Server, and \textit{CVE-2023-36884} again in the Windows 2019 Server on the Administrative Server. This illustrates how the same vulnerabilities can have different rankings depending on their deployment context and the criticality of the associated assets.

\section{Experimental Evaluation} \label{sec:Experiment}

We further evaluate our approach through systematic bench-marking against existing state-of-the-art methods. We did a systematic literature review of relevant works on vulnerability prioritization published after the year of 2013 in Scopus and IEEE Xplore database, resulting in 218 and 115 initial results from Scopus and IEEE Xplore, respectively. After merging and removing duplication and not peer-reviewed ones, there remained 265 papers. We then exclude papers that do not directly address cyber vulnerability prioritization or ranking, as well as papers not published in English. From the selected review papers, we further narrowed our focus to those that validate patch prioritization using real-world datasets and demonstrate their effectiveness in actual systems. Additionally, these papers must provide detailed descriptions of system configurations and topology or have published the corresponding configuration data. In doing so, we identified three application scenarios as detailed next.

\subsection{Scenario 1 - OpenPLC}

\subsubsection{Scenario Description}
\begin{figure}[h]
\centering
\includegraphics[width=0.36\textwidth]{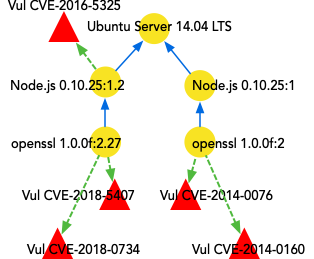}
\caption{Component configuration in Application Scenario 1.}
\label{fig:openPLC}
\end{figure}

\cite{longueira2022novel} provides a detailed description of an OpenPLC which is a fully open-source Programmable Logic Controller (PLC) designed for industrial IoT, and SCADA applications. OpenPLC normally consists of three main components: the runtime, which functions similarly to traditional PLC firmware; the editor, used for writing and compiling control programs; and the HMI builder, which creates web-based interfaces to monitor processes. The runtime is particularly important for its use of open-source libraries, such as libssl, and runs on various platforms, including Linux and Windows. In Scenario 1, the Server (Ubuntu Server) component has two Node.js components embedded in it. Each Node.js component has an OpenSSL library embedded within it, as illustrated in Fig. \ref{fig:openPLC}. Within this configuration, we examined several vulnerabilities affecting these components. For instance, OpenSSL 1.0.0f:2.27 is affected by \textit{CVE-2018-0734}, a cryptographic flaw that weakens encryption, and \textit{CVE-2018-5407}, a side-channel timing vulnerability that can expose sensitive information. \textit{CVE-2014-0160} impacts OpenSSL 1.0.0f:2 by allowing attackers to exploit buffer over-read issues, while \textit{CVE-2014-0076} is a cryptographic weakness that permits key recovery through side-channel attacks. Additionally, Node.js 0.10.25:1.2 is exposed to \textit{CVE-2016-5325}, a vulnerability allowing CRLF injection. 

\begin{table*}[h]
\centering
\scriptsize
\begin{tabular}{|c|c|c|c|c|c|c|c|c|c|}
\hline
Component & CVE & \begin{tabular}{@{}c@{}} CVSS \\V3 Base \end{tabular} & \begin{tabular}{@{}c@{}} Likelihood \\Subscore \end{tabular} & \begin{tabular}{@{}c@{}} Impact \\Subscore \end{tabular} & EPSS & \begin{tabular}{@{}c@{}} Exploit \\Exists \end{tabular} & \begin{tabular}{@{}c@{}} Scope \\Change \end{tabular} & \begin{tabular}{@{}c@{}} Ransom \\Utilized \end{tabular} 
& Rank \\ \hline
\multirow{2}{*}{\begin{tabular}{@{}c@{}} OpenSSL \\1.0.0f:2.27 \end{tabular}} 
& CVE-2018-0734  & 5.9 & 2.2 & 3.6 & 0.00342 & 0 & False & 0 & 3 \\ \cline{2-10}
& CVE-2018-5407  & 4.7 & 1.0 & 3.6 & 0.00061 & 0 & False & 0 & 4 \\ \hline

\multirow{2}{*}{\begin{tabular}{@{}c@{}} OpenSSL \\1.0.0f:2 \end{tabular}} 
& CVE-2014-0160 & 7.5 & 3.9 & 3.6 & 0.97354 & 1 & False & 0 & 2 \\ \cline{2-10} 
& CVE-2014-0076 & 2.8 & 1.3 & 1.4 & 0.00046 & 0 & False & 0 & 5 \\ \hline

Node.js & CVE-2016-5325  & 6.1 & 2.8 & 2.7 & 0.00437 & 0 & True & 0 & 1 \\ \hline
\end{tabular}
\vspace{5pt}
\caption{Vulnerabilities and Component Information in Application Scenario 1}
\label{tab:AS1vulnerabilities}
\end{table*}

\subsubsection{Result}

Our vulnerability assessment and patch prioritization rank is listed in Table \ref{tab:AS1vulnerabilities}, which is aligned with the comment in \cite{longueira2022novel}: \textquotedblleft \textit{We could argue that, as Node.js is the most accessible asset from the exterior, its vulnerabilities should be first addressed, even though the associated CVSS is not the highest one.}\textquotedblright

\subsection{Scenario 2 - Network Environment}

\subsubsection{Scenario Description}
In Scenario 2, the analysis focuses on a network environment consisting of six key assets, including firewalls, VPN appliances, mail servers, web servers, and a domain controller, modified from the case study scenario in \cite{zeng2023illation}. A preliminary description of the network configuration is provided in \cite{zeng2023illation}, but it was incomplete with limited information on host configuration. Furthermore, the unexploited vulnerabilities selected in \cite{zeng2023illation} are not well-aligned with the vulnerable component configuration found in the exploited vulnerabilities set. For instance, the exploited vulnerability \textit{CVE-2020-5902} is typically associated with the F5 Networks firewall, which would not ordinarily be exposed to the vulnerability \textit{CVE-2020-7995}. To enhance the configuration of Scenario 2, we conducted in-depth research into CISA reports \cite{cisa-apt} related to the relevant APT attacks and the CPE configurations of the confirmed existing vulnerabilities. 

\begin{figure}[h]
\centering
\includegraphics[width=0.47\textwidth]{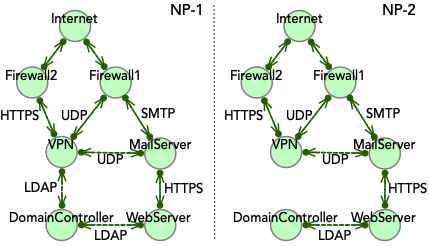}
\caption{Host reachability in Application Scenario 2.}
\label{fig:IllationRepro}
\end{figure}

The scenario distinguishes between two network configurations: NP-1 and NP-2. NP-1 and NP-2 differ by whether the HTTPS connection between VPN and DomainController is blocked. This adjustment affects the network's reachability and alters the prioritization of vulnerabilities based on the new network constraints, as illustrated in Fig. \ref{fig:IllationRepro}.

\begin{table*}[h]
\centering
\scriptsize
\begin{tabular}{|c|c|c|c|c|c|c|c|c|c|c|c|}
\hline
\textbf{Asset} & \begin{tabular}{@{}c@{}} Criticality \\(NP-1; NP-2) \end{tabular}
& CVE & \begin{tabular}{@{}c@{}} CVSS \\V3 Base \end{tabular}
& \begin{tabular}{@{}c@{}} Likelihood \\Subscore \end{tabular}
& \begin{tabular}{@{}c@{}} Impact \\Subscore \end{tabular}
& EPSS
& \begin{tabular}{@{}c@{}} Exploit \\Exists \end{tabular}
& \begin{tabular}{@{}c@{}} Scope \\Change \end{tabular}
&\begin{tabular}{@{}c@{}} Ransom \\Utilized \end{tabular}  
& \begin{tabular}{@{}c@{}}  Rank \\ NP-1 \end{tabular}
& \begin{tabular}{@{}c@{}}  Rank \\ NP-2 \end{tabular}
\\ \hline
\multirow{1}{*}{Firewall1} & 0.45; 0.53 & CVE-2020-5902 & 9.8 & 3.9 & 5.9 & 0.97542 & 1 & False & 1 & 3 & 4 \\ \hline
\multirow{2}{*}{\begin{tabular}{@{}c@{}} VPN \\Appliance \end{tabular}} & \multirow{2}{*}{0.47; 0.46} & CVE-2019-11510 & 10.0 & 6.0 & 3.9 & 0.97297 & 1 & True & 1 & 2 & 7 \\ \cline{3-12} 
& & CVE-2019-11540 & 8.3 & 1.6 & 6.0 & 0.03547 & 0 & True & 0 & 10 & 10 \\ \hline
\multirow{2}{*}{\begin{tabular}{@{}c@{}} Mail \\Server \end{tabular}} & \multirow{2}{*}{0.56; 0.64} & CVE-2019-19781 & 9.8 & 3.9 & 5.9 & 0.97538 & 1 & False & 1 & 5 & 1 \\ \cline{3-12} 
 &  & CVE-2022-27518 & 9.8 & 3.9 & 5.9 & 0.20483 & 1 & True & 0 & 7 & 2 \\ \hline
\multirow{1}{*}{Firewall2} & 0.34; 0.30 & CVE-2020-1631 & 9.8 & 3.9 & 5.9 & 0.00529 & 1 & False & 0 & 8 & 9 \\ \hline
\multirow{2}{*}{\begin{tabular}{@{}c@{}} Web \\Server \end{tabular}} & \multirow{2}{*}{0.48; 0.59} & CVE-2018-13379 & 9.8 & 3.9 & 5.2 & 0.96863 & 1 & False & 1 & 6 & 5 \\ \cline{3-12} 
 &  & CVE-2021-43206 & 4.3 & 2.8 & 1.4 & 0.00072 & 0 & False & 0 & 11 & 11 \\ \hline
\multirow{3}{*}{\begin{tabular}{@{}c@{}} Domain \\Controller \end{tabular}} & \multirow{3}{*}{0.63; 0.66} & CVE-2020-1472 & 10.0 & 3.9 & 6.0 & 0.40299 & 1 & True & 1 & 1 & 3 \\ \cline{3-12} 
 &  & CVE-2021-31955 & 5.5 & 1.8 & 3.6 & 0.01042 & 1 & False & 0 & 9 & 8 \\ \cline{3-12} 
 &  & CVE-2021-42278 & 8.8 & 2.8 & 5.9 & 0.00747 & 1 & False & 1 & 4 & 6 \\ \hline
\end{tabular}
\vspace{5pt}
\caption{Vulnerabilities and Asset Information in Application Scenario 2}
\label{tab:AS2vulnerabilities}
\end{table*}

\subsubsection{Result}
Our patch ranking is listed in Table \ref{tab:AS2vulnerabilities}, and reflects the variations in vulnerability prioritization between the NP-1 and NP-2 network scenarios. The rankings remain consistent for some critical vulnerabilities, yet notable shifts occur due to the structural changes in the network.

In NP-1, the Domain Controller's vulnerabilities, particularly \textit{CVE-2020-1472}, are ranked highest due to its direct interaction with the VPN. This connection significantly increases the potential impact and likelihood of an exploit, making it a top priority for patching. However, in NP-2, the removal of the communication link between the VPN and the Domain Controller results in a slight reduction in the criticality of these vulnerabilities, causing \textit{CVE-2020-1472} to drop from the top position to third place. This shift highlights how network architecture changes can influence vulnerability prioritization, as the decreased interaction reduces the exposure and potential attack vectors against the Domain Controller.

Interestingly, the criticality scores of other assets, such as the Mail Server and Firewall1, exhibit an increase in NP-2. This adjustment emphasizes the importance of these assets in the absence of the VPN-DC link. Specifically, vulnerabilities like \textit{CVE-2019-19781} in the Mail Server move to the top of the priority list in NP-2, underscoring its heightened risk in the revised network configuration.

Our analysis also reveals that the removal of the VPN-DC connection in NP-2 affects the vulnerability ranking of the Web Server. The vulnerability \textit{CVE-2018-13379} on the Web Server remains critical but shifts from sixth to fifth place in NP-2, reflecting the changing risk dynamics as the network's dependency on the VPN diminishes.

We partially implemented the methodology of \cite{zeng2023illation} and \cite{zeng2021licality} on our modified dataset, utilizing our EPSS scores as the initial assessment score in the ILLATION framework due to the unavailability of their specific model for initial score calculation. For NP-1, the top vulnerabilities identified using \cite{zeng2021licality} are \textit{CVE-2018-13379} (0.4649), \textit{CVE-2019-11510} (0.4573), and \textit{CVE-2020-5902} (0.4389). In NP-2, the same vulnerabilities remain prioritized, though with increased scores: \textit{CVE-2018-13379} (0.5715), \textit{CVE-2020-5902} (0.5170), and \textit{CVE-2019-11510} (0.4476).


In comparison, our approach offers a ranking that accurately reflects the changes in network architecture, highlighting critical vulnerabilities such as \textit{CVE-2020-1472} in NP-1 and its reduced impact in NP-2. This adaptability in our approach to capture the effects of network structural changes on vulnerability prioritization is crucial for effective vulnerability management and network security.

\subsection{Scenario 3 - Industrial Control System}

\subsubsection{Scenario Description}

Application Scenario 3 involves a structured SCADA network with three foundational layers: the enterprise, industrial, and field layers, adapted from the case study in \cite{cheimonidis2024dynamic}. The enterprise layer is secured by a firewall and includes key assets like the VPN for remote access, the Web Server (WebS) managing the company’s website, and Workstations (WS) facilitating communication with the industrial layer. These components are vulnerable to vulnerabilities, such as \textit{CVE-2019-11510} affecting the VPN, which allows unauthorized file access, and \textit{CVE-2017-7269} impacting the WebS, enabling remote code execution. The WS is also at risk from \textit{CVE-2017-0143} and \textit{CVE-2017-8692}, which pose further remote code execution threats. In the industrial layer, the Historian Database (HDB), Human Machine Interface (HMI), and Engineering Workstation (EWS) are critical for SCADA infrastructure management. The HDB is vulnerable to \textit{CVE-2021-1636} and \textit{CVE-2023-21528}, allowing potential unauthorized access and code execution. The HMI, targeted by \textit{CVE-2016-5743}, and the EWS, vulnerable to \textit{CVE-2019-10922}, present risks for remote code execution. Finally, in the field layer, PLC, responsible for controlling physical operations, is exposed to \textit{CVE-2016-9159} and \textit{CVE-2016-8673}, which could compromise its credentials or authentication, leading to potential operational disruptions. In terms of network connectivity, the VPN and WebS are directly connected to the WS, which communicates with industrial layer assets such as HDB, HMI, and EWS. The PLC, in turn, is directly connected to HDB, HMI, and EWS.

\subsubsection{Result}

Table \ref{tab:AS3vulnerabilities} presents the ranking of vulnerabilities across different assets in the enterprise network. Notably, the top vulnerabilities identified using our model vary from the rankings produced by CVSS scores. For instance, vulnerabilities within the Workstation \textit{CVE-2017-0143} and \textit{CVE-2017-8692} and the VPN Server \textit{CVE-2019-11510} are highly ranked due to factors like exploitability and ransomware utilization, which play a significant role in our assessment. 

\begin{table*}[h]
\centering
\scriptsize
\begin{tabular}{|c|c|c|c|c|c|c|c|c|c|c|}
\hline
\textbf{Asset} & \textbf{Criticality} & \textbf{CVE} & \begin{tabular}{@{}c@{}} CVSS \\ V3 Base \end{tabular} & \begin{tabular}{@{}c@{}} Likelihood \\ Subscore \end{tabular} & \begin{tabular}{@{}c@{}} Impact \\ Subscore \end{tabular} & EPSS & \begin{tabular}{@{}c@{}} Exploit \\ Exists \end{tabular} & \begin{tabular}{@{}c@{}} Scope \\ Change \end{tabular} & \begin{tabular}{@{}c@{}} Ransom \\ Utilized \end{tabular} & \begin{tabular}{@{}c@{}} \textbf{Rank} \end{tabular} \\ \hline

\multirow{2}{*}{Workstation} & \multirow{2}{*}{0.55} & CVE-2017-0143 & 8.1 & 2.2 & 5.9 & 0.96304 & 1 & False & 1 & 1 \\ \cline{3-11} 
 &  & CVE-2017-8692 & 7.5 & 1.6 & 5.9 & 0.14441 & 1 & False & 0 & 2 \\ \hline

VPN Server & 0.47 & CVE-2019-11510 & 10.0 & 3.9 & 6.0 & 0.97297 & 1 & True & 1 & 3 \\ \hline

Web Server & 0.80 & CVE-2017-7269 & 9.8 & 3.9 & 5.9 & 0.97321 & 0 & False & 0 & 4 \\ \hline

\multirow{2}{*}{PLC} & \multirow{2}{*}{0.50} & CVE-2016-8673 & 8.8 & 2.8 & 5.9 & 0.00108 & 0 & False & 0 & 5 \\ \cline{3-11}  
 &  & CVE-2016-9159 & 5.9 & 2.2 & 3.6 & 0.002 & 0 & False & 0 & 9 \\ \hline

HMI & 0.58 & CVE-2016-5743 & 9.8 & 3.9 & 5.9 & 0.04346 & 0 & False & 0 & 6 \\ \hline

\multirow{2}{*}{\begin{tabular}{@{}c@{}} Historian \\ Database \end{tabular}} & \multirow{2}{*}{0.58} & CVE-2021-1636 & 8.8 & 2.8 & 5.9 & 0.00157 & 0 & False & 0 & 7 \\ \cline{3-11}  
 &  & CVE-2023-21528 & 7.8 & 1.8 & 5.9 & 0.00081 & 0 & False & 0 & 10 \\ \hline

\begin{tabular}{@{}c@{}} Engineering \\ Workstation \end{tabular} & 0.65 & CVE-2019-10922 & 9.8 & 3.9 & 5.9 & 0.00342 & 0 & False & 0 & 8 \\ \hline

\end{tabular}
\vspace{5pt}
\caption{Vulnerabilities and Asset Information in Application Scenario 3}
\label{tab:AS3vulnerabilities}
\end{table*}

\cite{cheimonidis2024dynamic} emphasizes the importance of addressing vulnerabilities within the PLC and WS (Workstation) assets first, given the unique challenges of patching in SCADA environments. Similarly, our presented Table \ref{tab:AS3vulnerabilities} also prioritizes vulnerabilities in the Workstation and VPN Server assets, recognizing their high impact, likelihood, and the fact that some have been exploited by ransomware. However, vulnerabilities in the PLC and HMI assets, while still important, are ranked lower due to practical considerations such as the feasibility of patching in real-world environments.

\section{Discussion} \label{sec:Discussion}

Our case study in Section \ref{sec:CaseStudy} illustrates the complex challenges security teams face in prioritizing vulnerabilities within resource constraints while minimizing operational disruption. The diverse vulnerability landscape, encompassing web server exposures, critical remote code execution vulnerabilities on application servers, and threats propagating across user workstations, necessitates a systematic risk assessment approach.

Our methodology addresses these challenges through a multi-metric scoring model integrated with functional dependency analysis. This model assesses asset criticality and calculates both direct and indirect risks associated with each vulnerability. By incorporating factors such as exploitability, potential impact, and asset interdependencies, our approach enables security teams to optimize patching efforts for maximum overall network risk reduction.

In our experimental evaluation in Section \ref{sec:Experiment} across the three distinct application scenarios, we leveraged four key metrics to assess the effectiveness of our vulnerability prioritization approach: explainability, granularity and generalizability.

\begin{itemize}
    \item Explainability refers to how well a risk analysis method conveys its results, making them understandable and actionable for stakeholders. 
    \item Granularity denotes the level of detail in the analysis, from high-level overviews (coarse-grained) to in-depth examinations of individual vulnerabilities and attack paths (fine-grained).
    \item Generalizability measures the ability of a method to be applied effectively across a wide range of systems, environments, and industries, reflecting its flexibility and capacity to handle diverse configurations.
\end{itemize}

Our approach excels in explainability by providing clear, context-aware rankings of vulnerabilities. The rankings reflect the real-world implications of each vulnerability, considering factors such as exploitability, the presence of ransomware, and the criticality of the affected assets. For example, in Scenario 3, the vulnerabilities within the Workstation are prioritized due to their high likelihood of exploitation and ransomware utilization, despite their moderate CVSS scores. This level of contextual insight ensures that security engineers can make informed decisions and prioritize patches that address the most pressing threats to the system.

While traditional methods often provide a high-level overview, our approach offers both coarse-grained and fine-grained analysis, considering not just the CVSS base scores but also sub-components such as likelihood, impact, and the presence of exploit. In Scenario 1, for instance, the vulnerability within the Node.js component is prioritized over those in the OpenSSL libraries, despite the latter having higher CVSS scores. Our approach also demonstrates a high degree of adaptability, particularly in scenarios where network architecture changes impact vulnerability prioritization. In Scenario 2, the shift in vulnerability rankings between NP-1 and NP-2 configurations highlights our method's ability to dynamically adjust to new security landscapes. This contrasts with traditional CVSS-based methods, which may not fully account for the variation introduced by network modifications.

Generalizability is reflected in our method's applicability across different scenarios, from industrial control systems to enterprise networks. The approach was effectively applied to diverse environments in Scenarios 1, 2, and 3, each with its own unique configurations. Scenario 1 introduces a typical enterprise Information Technology (IT) network with a mix of server and client systems. Scenario 2 involves changes in network topology to evaluate the method’s performance under varying configurations. Scenario 3 focuses on an Industrial Control System (ICS) environment, characterized by a unique set of dependencies and risk factors distinct from traditional IT environments. This generalizability is underpinned by the hierarchical structure of our system graph, $\mathcal{G}$, which facilitates the modeling of varied system architectures and asset attributes distributed across multiple domains. Furthermore, our dependence rules framework incorporates functional dependencies prevalent in both IT and ICS environments. These dependencies are differentiated through the assignment of empirically derived weights, reflecting the practical implications of each relationship. This approach ensures that our model accurately captures the inter-connected characteristic of complex, multi-domain systems.

\section{Conclusion} \label{sec:Conclusion}

This paper introduces a graph-based framework for vulnerability prioritization and risk aggregation, designed to address the complexities of modern systems. Our approach extends beyond traditional methods by integrating detailed asset and component information to generate precise CPE 2.3 metadata, which is used to query vulnerability databases (e.g., NVD, Shodan, etc.) for applicable vulnerabilities and threat intelligence. The proposed dual-graph model, comprising of a network communication graph and a system dependency graph, captures potential attack paths and the cascading effects of vulnerabilities across system components. This holistic perspective is crucial for understanding risk propagation through interconnected systems, enabling more strategic security decisions. Our framework prioritizes vulnerabilities based on exploit likelihood, impact, and risk propagation, optimizing resource allocation for maximum system security.

Through benchmarking against state-of-the-art methods, our framework has demonstrated accuracy, explainability and adaptability. It effectively addresses the limitations of traditional CVSS-based approaches by considering the specific context of each vulnerability, including the operational environment and the criticality of affected assets. This context-driven approach not only improves the precision of vulnerability prioritization, but also enhances the generalizability of our framework across diverse systems and industries.

Future research will focus on further refining our graph-based models to incorporate real-time threat intelligence and machine learning techniques for predictive analytics. Additionally, we plan to expand the application of our framework to different industry sectors, including critical infrastructure and IoT environments, to assess its scalability and robustness in varying operational contexts.

\bibliographystyle{ACM-Reference-Format}
\bibliography{VulRG}

\appendix

\section{Appendix: Notation}

Table \ref{tab:notations} summarizes the notations used in this paper for risk analytics.

\begin{table*}[h]
\centering
\scriptsize
\caption{Definitions of Notations}
\begin{tabular}{|l|l|}
\hline
\textbf{Notation} & \textbf{Definition} \\ \hline
\multicolumn{2}{|l|}{\textbf{Indices:}} \\ \hline
$\mathcal{S}$ & System graph \\ \hline
$H$ & Set of hosts in the system \\ \hline
$\mathcal{A}$ & Set of assets \\ \hline
$\mathcal{A}_h$ & Set of high-criticality assets \\ \hline
$h_i$ & Singular host node in the system \\ \hline
$a_j$ & Singular asset node in the system \\ \hline
$C$ & Set of components \\ \hline
$c_i$ & Singular component node \\ \hline
$V(c_i)$ & Set of vulnerabilities associated with component $c_i$ \\ \hline
$v_k$ & Singular vulnerability \\ \hline
$\mathcal{E}$ & Set of edges in the system graph \\ \hline
$\mathcal{E}_c$ & Edges indicating network communication links between assets \\ \hline
$\mathcal{E}_f$ & Edges denoting functional dependencies between components across assets \\ \hline
$E$ & Set of edges in the asset sub-graph \\ \hline
$E_f$ & Edges denoting functional dependencies between components \\ \hline
$E_e$ & Edges denoting embedding relationships between components \\ \hline
\multicolumn{2}{|l|}{\textbf{Attributes:}} \\ \hline
$\text{IP}(a_k)$ & IP address of asset $a_k$ \\ \hline
$\text{MAC}(a_k)$ & MAC address of asset $a_k$ \\ \hline
$\text{Subnet}(a_k)$ & Subnet within which the asset $a_k$ resides \\ \hline
$vendor(c_i)$ & Vendor of component $c_i$ \\ \hline
$name(c_i)$ & Name of component $c_i$ \\ \hline
$version(c_i)$ & Version of component $c_i$ \\ \hline
$S^{\text{cvss}}_v$ & CVSS score of vulnerability $v$ \\ \hline
$S^{\text{likelihood}}_v$ & Likelihood score of vulnerability $v$ \\ \hline
$S^{\text{impact}}_v$ & Impact score of vulnerability $v$ \\ \hline
$\text{Exploit}(v_{c_i}^k)$ & Exploit indicator of vulnerability $v_{c_i}^k$ \\ \hline
$\text{ScopeChange}(v_k)$ & Scope change indicator for vulnerability $v_k$ \\ \hline
$\text{Ransomware}(v_k)$ & Historical ransomware use indicator for vulnerability $v_k$ \\ \hline
\multicolumn{2}{|l|}{\textbf{Weights:}} \\ \hline
$w_c(a_i, a_j)$ & Communication weight between assets $a_i$ and $a_j$ \\ \hline
$w_f(c_i, c_j)$ & Functional dependency weight between components $c_i$ and $c_j$ \\ \hline
$w_e(c_i, c_j)$ & Embedding relationship weight between components $c_i$ and $c_j$ \\ \hline
$w_l$ & Weighting factor for severity level $l$ in CVS \\ \hline
$\alpha$ & Weight for CVSS likelihood score in exploit likelihood \\ \hline
$\beta$ & Weight for EPSS score in exploit likelihood \\ \hline
$\gamma$ & Weight for exploit existence in exploit likelihood \\ \hline
$\delta$ & Weight for scope change indicator in propagation likelihood \\ \hline
$\theta$ & Weight for ransomware use indicator in propagation likelihood \\ \hline
$w_1$ and $w_2$ & Weights for the relative importance of centrality and business criticality \\ \hline
\multicolumn{2}{|l|}{\textbf{Risk Metrics:}} \\ \hline
$\mathbf{R}_{\text{component}, c_i}$ & Risk level of component $c_i$ \\ \hline
$\mathbf{R}_{\text{component}, c_i}^{\text{CVS}}$ & Component Vulnerability Score (CVS) of $c_i$ \\ \hline
$\mathbf{R}_{\text{component}, c_i}^{\text{Centrality}}$ & Centrality of component $c_i$ \\ \hline
$\mathbf{R}_{\text{asset}, a_m}$ & Overall risk score for asset $a_m$ \\ \hline
$\mathbf{R}_{\text{vulnerability}, v_k}$ & Risk associated with vulnerability $v_k$ \\ \hline
$\mathbf{R}_{\text{direct}, v_k}$ & Direct vulnerability risk of $v_k$ \\ \hline
$\mathbf{R}_{\text{indirect}, v_k}$ & Indirect vulnerability risk of $v_k$ \\ \hline
$\mathbf{R}_{\text{network}}$ & Network-based risk for the entire system \\ \hline
$\mathbf{R}_{\text{host}, h_m}$ & Host-based risk for host $h_m$ \\ \hline
$\mathbf{R}_{\text{system}}$ & Total system-level risk \\ \hline
$\mathbf{R}_{\text{asset}, a_m}^{\text{Criticality}}$ & Criticality of asset $a_m$ \\ \hline
$\mathbf{R}_{\text{asset}, a_m}^{\text{BusinessCriticality}}$ & Business criticality of asset $a_m$ \\ \hline
$\mathbf{R}_{\text{asset}, a_m}^{\text{Centrality}}$ & Centrality of asset $a_m$ based on its components \\ \hline
$\text{SAP}_{(a_s, a_h, \text{path})}$ & Shortest attack path from source $a_s$ to high-criticality asset $a_h$ \\ \hline
\end{tabular}
\label{tab:notations}
\end{table*}

\clearpage

\section{Appendix: Search Strings}

Table \ref{tab:SearchStrings} presents the search strings used to query Scopus and IEEE Xplore for relevant literature.

\begin{table*} [h]
\begin{centering}
\small
  \caption{Search Strings for the Review}
  \label{tab:SearchStrings}
  \resizebox{1\textwidth}{!}{%
  \begin{tabularx}{\textwidth}{|l|X|}
    \hline
    Database & Search string
    \\
    \hline
    IEEE Xplore & \textit{(("Full Text \& Metadata":"vulnerability prioriti*" OR "Full Text \& Metadata":"patch rank*" OR "Full Text \& Metadata":"vulnerability rank*" OR "Full Text \& Metadata":"patch prioriti*" OR "Full Text \& Metadata":"risk aggregat*" OR "Full Text \& Metadata":"vulnerability aggregat*") AND ("Full Text \& Metadata":"cyber security" OR "Full Text \& Metadata":"cybersecurity"))} 
    \\
    \hline
    Scopus & \textit{(("vulnerability prioriti*" OR "patch rank*" OR "vulnerability rank*" OR "risk aggregat*" OR "patch prioriti*" OR "vulnerability aggregat*") AND ("cybersecurity" OR "cyber security") AND PUBYEAR $\ge$ 2013)}
    \\
    \hline
\end{tabularx}}
\end{centering}
\end{table*}

\section{Appendix: Vulnerabilities and Asset Information}

Tables \ref{tab:completeRank_I} and \ref{tab:completeRank_II} presents the detailed information of vulnerabilities and asset information for the enterprise network system used in our case study, detailing each component's version, associated CVEs, CVSS V3 Base scores, likelihood and impact subscores, EPSS values, exploit existence, scope change indicators, ransomware utilization, and patch ranking.

\begin{table*}[h]
\centering
\scriptsize
\caption{Vulnerabilities and Asset Information in Enterprise Network (Part I)}
\rotatebox{90}{
\begin{tabular}{|c|c|c|c|c|c|c|c|c|c|c|c|}
\hline
\textbf{Asset} & \begin{tabular}{@{}c@{}} Asset \\Criticality \end{tabular}
& \textbf{Component} & CVE & \begin{tabular}{@{}c@{}} CVSS \\V3 Base \end{tabular}
& \begin{tabular}{@{}c@{}} Likelihood \\Subscore \end{tabular}
& \begin{tabular}{@{}c@{}} Impact \\Subscore \end{tabular}
& EPSS
& \begin{tabular}{@{}c@{}} Exploit \\Exists \end{tabular}
& \begin{tabular}{@{}c@{}} Scope \\Change \end{tabular}
& \begin{tabular}{@{}c@{}} Ransomware \\Utilized \end{tabular}  
& \begin{tabular}{@{}c@{}} Patch \\Rank \end{tabular}
\\ \hline

\multirow{9}{*}{\begin{tabular}{@{}c@{}} Web \\Server \end{tabular}} 
& \multirow{9}{*}{2} 
& \multirow{3}{*}{\begin{tabular}{@{}c@{}} Ubuntu \\(20.04) \end{tabular}} 
& CVE-2022-0492 & 7.8 & 1.8 & 5.9 & 0.09515 & 1 & False & 0 & 24 \\ \cline{4-12} 
&  &  & CVE-2020-25719 & 7.2 & 1.2 & 5.9 & 0.00123 & 0 & False & 0 & 41 \\ \cline{4-12} 
&  &  & CVE-2021-3156 & 7.8 & 1.8 & 5.9 & 0.96133 & 1 & False & 0 & 12 \\ \cline{3-12}

& & Apache HTTP (2.4.49) & CVE-2021-41773 & 7.5 & 3.9 & 3.6 & 0.97447 & 1 & False & 1 & 7 \\ \cline{3-12}

& & \multirow{5}{*}{\begin{tabular}{@{}c@{}} Apache Tomcat \\(9.0.2) \end{tabular}} 
& CVE-2020-1938 & 9.8 & 3.9 & 5.9 & 0.97284 & 1 & False & 0 & 9 \\ \cline{4-12} 
&  &  & CVE-2024-21733 & 5.3 & 3.9 & 1.4 & 0.00692 & 0 & False & 0 & 48 \\ \cline{4-12} 
&  & & CVE-2023-46589 & 7.5 & 3.9 & 3.6 & 0.00484 & 0 & False & 0 & 30 \\ \cline{4-12} 
 &  & &CVE-2023-41080 & 6.1 & 2.8 & 2.7 & 0.00244 & 0 & True & 0 & 42 \\ \cline{4-12} 
 &  & & CVE-2019-0221 & 6.1 & 2.8 & 2.7 & 0.01096 & 1 & True & 0 & 26 \\ \hline

\multirow{6}{*}{\begin{tabular}{@{}c@{}} DNS \\Server \end{tabular}} 
& \multirow{6}{*}{2} 
& \multirow{3}{*}{\begin{tabular}{@{}c@{}} Debian Linux \\(11.0) \end{tabular}} 
& CVE-2023-51385 & 6.5 & 3.9 & 2.5 & 0.00267 & 0 & False & 0 & 49 \\ \cline{4-12} 
&  &  & CVE-2023-44487 & 7.5 & 3.9 & 3.6 & 0.81041 & 1 & False & 0 & 25 \\ \cline{4-12} 
&  &  & CVE-2023-43641 & 8.8 & 2.8 & 5.9 & 0.00136 & 0 & False & 0 & 37 \\ \cline{3-12}

& & \multirow{2}{*}{\begin{tabular}{@{}c@{}} ISC BIND \\(9.16.1) \end{tabular}} 
& CVE-2020-8625 & 8.1 & 2.2 & 5.9 & 0.18732 & 0 & False & 0 & 44 \\ \cline{4-12} 
&  &  & CVE-2020-8616 & 8.6 & 3.9 & 4.0 & 0.02336 & 0 & True & 0 & 52 \\ \cline{3-12}

&  & PowerDNS Recursor (4.5.3) & CVE-2021-25215 & 7.5 & 3.9 & 3.6 & 0.0674 & 0 & False & 0 & 51 \\ \hline

\multirow{8}{*}{\begin{tabular}{@{}c@{}} Application \\Server \end{tabular}} 
& \multirow{8}{*}{5} 
& \multirow{2}{*}{\begin{tabular}{@{}c@{}} Debian Linux \\(12.0) \end{tabular}} & CVE-2024-6387 & 8.1 & 2.2 & 5.9 & 0.71475 & 0 & False & 0 & 10 \\ \cline{4-12}
&  &  & CVE-2023-47272 & 6.1 & 2.7 & 2.8 & 0.00098 & 0 & True & 0 & 2 \\ \cline{3-12}

& & \multirow{3}{*}{\begin{tabular}{@{}c@{}} Oracle WebLogic Server \\(12.2.1.4.0) \end{tabular}}  & CVE-2023-21839 & 7.5 & 3.9 & 3.6 & 0.95721 & 1 & False & 0 & 6 \\ \cline{4-12}
&  &  & CVE-2022-22965 & 7.5 & 3.9 & 3.6 & 0.97485 & 1 & False & 0 & 5 \\ \cline{4-12}
&  &  & CVE-2024-20986 & 6.1 & 2.8 & 2.7 & 0.00046 & 0 & True & 0 & 14 \\ \cline{3-12}

& & \multirow{2}{*}{\begin{tabular}{@{}c@{}} Java SE Runtime Environment \\(8u281) \end{tabular}}  & CVE-2021-2163 & 5.3 & 1.6 & 3.6 & 0.00224 & 0 & False & 0 & 43 \\ \cline{4-12}
&  &  & CVE-2021-2161 & 5.9 & 2.2 & 3.6 & 0.00353 & 0 & False & 0 & 35 \\ \cline{3-12}

&& Apache HTTP (2.4.49)  & CVE-2021-41773 & 7.5 & 3.9 & 3.6 & 0.97447 & 1 & False & 1 & 11 \\ \hline

\multirow{7}{*}{\begin{tabular}{@{}c@{}} Database \\Server \end{tabular}} 
& \multirow{7}{*}{5} 
& \multirow{4}{*}{\begin{tabular}{@{}c@{}} Windows Server \\(2019) \end{tabular}} 
& CVE-2021-26414 & 6.5 & 3.6 & 2.8 & 0.0005 & 0 & False & 0 & 34 \\ \cline{4-12} 
&  &  & CVE-2024-21338 & 7.8 & 1.8 & 5.9 & 0.01426 & 1 & False & 0 & 15 \\ \cline{4-12} 
&  &  & CVE-2023-36884 & 7.5 & 1.6 & 5.9 & 0.09991 & 1 & False & 1 & 1 \\ \cline{4-12} 
&  &  & CVE-2023-36028 & 9.8 & 3.9 & 5.9 & 0.57012 & 0 & False & 0 & 16 \\ \cline{3-12}

& & \multirow{2}{*}{\begin{tabular}{@{}c@{}} SQL Server \\(2019) \end{tabular}} 
& CVE-2023-38169 & 8.8 & 2.8 & 5.9 & 0.02826 & 0 & False & 0 & 38 \\ \cline{4-12} 
&  &  & CVE-2023-36728 & 5.5 & 1.8 & 3.6 & 0.00049 & 0 & False & 0 & 54 \\ \cline{3-12}

& & Veeam Backup (11a) & CVE-2021-35971 & 9.8 & 3.9 & 5.9 & 0.00683 & 0 & False & 0 & 33 \\ \hline

\multirow{6}{*}{\begin{tabular}{@{}c@{}} FTP \\Server \end{tabular}} 
& \multirow{6}{*}{3} 
& \multirow{3}{*}{\begin{tabular}{@{}c@{}} Ubuntu Server \\(22.04) \end{tabular}} 
& CVE-2022-34918 & 7.8 & 1.8 & 5.9 & 0.00635 & 1 & False & 0 & 22 \\ \cline{4-12} 
&  &  & CVE-2023-1523 & 10.0 & 3.9 & 6.0 & 0.0007 & 0 & True & 0 & 8 \\ \cline{4-12} 
&  &  & CVE-2022-29581 & 7.8 & 1.8 & 5.9 & 0.00042 & 0 & False & 0 & 40 \\ \cline{3-12}

& & \multirow{2}{*}{\begin{tabular}{@{}c@{}} FileZilla Server \\(4.3.3) \end{tabular}} 
& CVE-2020-26299 & 9.6 & 3.1 & 5.8 & 0.00337 & 0 & True & 0 & 32 \\ \cline{4-12} 
&  &  & CVE-2020-15152 & 9.1 & 3.9 & 5.2 & 0.0032 & 0 & False & 0 & 31 \\ \hline

\end{tabular}}
\label{tab:completeRank_I}
\end{table*}

\begin{table*}[h]
\centering
\scriptsize
\caption{Vulnerabilities and Asset Information in Enterprise Network (Part II)}
\rotatebox{90}{
\begin{tabular}{|c|c|c|c|c|c|c|c|c|c|c|c|}
\hline
\textbf{Asset} & \begin{tabular}{@{}c@{}} Asset \\Criticality \end{tabular}
& \textbf{Component} & CVE & \begin{tabular}{@{}c@{}} CVSS \\V3 Base \end{tabular}
& \begin{tabular}{@{}c@{}} Likelihood \\Subscore \end{tabular}
& \begin{tabular}{@{}c@{}} Impact \\Subscore \end{tabular}
& EPSS
& \begin{tabular}{@{}c@{}} Exploit \\Exists \end{tabular}
& \begin{tabular}{@{}c@{}} Scope \\Change \end{tabular}
& \begin{tabular}{@{}c@{}} Ransomware \\Utilized \end{tabular}  
& \begin{tabular}{@{}c@{}} Patch \\Rank \end{tabular}
\\ \hline
\multirow{10}{*}{\begin{tabular}{@{}c@{}} Administrative \\Server \end{tabular}} 
& \multirow{10}{*}{4} 
& \multirow{4}{*}{\begin{tabular}{@{}c@{}} Windows Server \\(2019) \end{tabular}} 
& CVE-2021-26414 & 6.5 & 3.6 & 2.8 & 0.0005 & 0 & False & 0 & 39 \\ \cline{4-12} 
&  &  & CVE-2024-21338 & 7.8 & 1.8 & 5.9 & 0.01426 & 1 & False & 0 & 17 \\ \cline{4-12} 
&  &  & CVE-2023-36884 & 7.5 & 1.6 & 5.9 & 0.09991 & 1 & False & 1 & 3 \\ \cline{4-12} 
&  &  & CVE-2023-36028 & 9.8 & 3.9 & 5.9 & 0.57012 & 0 & False & 0 & 19 \\ \cline{3-12}

& & \multirow{3}{*}{\begin{tabular}{@{}c@{}} Nagios XI \\(5.8.5) \end{tabular}} 
& CVE-2021-25296 & 8.8 & 2.8 & 5.9 & 0.96507 & 1 & False & 0 & 18 \\ \cline{4-12} 
&  &  & CVE-2023-51072 & 5.4 & 2.3 & 2.7 & 0.00095 & 0 & True & 0 & 55 \\ \cline{4-12} 
&  &  & CVE-2023-40933 & 8.8 & 2.8 & 5.9 & 0.00097 & 0 & False & 0 & 45 \\ \cline{3-12}

& & \multirow{3}{*}{\begin{tabular}{@{}c@{}} Splunk Enterprise \\(8.1.0) \end{tabular}} 
& CVE-2023-32707 & 8.8 & 2.8 & 5.9 & 0.89852 & 1 & False & 0 & 20 \\ \cline{4-12} 
&  &  & CVE-2023-40598 & 8.5 & 1.8 & 6.0 & 0.00065 & 0 & True & 0 & 53 \\ \cline{4-12} 
&  &  & CVE-2023-32706 & 7.7 & 3.1 & 4.0 & 0.00055 & 0 & True & 0 & 50 \\ \hline

\multirow{9}{*}{\begin{tabular}{@{}c@{}} User \\Workstation \end{tabular}} 
& \multirow{9}{*}{3} 
& \multirow{3}{*}{\begin{tabular}{@{}c@{}} Windows 10 \\(10.0.19044.3448) \end{tabular}} 
& CVE-2024-30051 & 7.8 & 1.8 & 5.9 & 0.00048 & 1 & False & 0 & 23 \\ \cline{4-12} 
&  &  & CVE-2024-29988 & 8.8 & 2.8 & 5.9 & 0.00226 & 1 & False & 0 & 21 \\ \cline{4-12} 
&  &  & CVE-2024-26169 & 7.8 & 1.8 & 5.9 & 0.00316 & 1 & False & 1 & 4 \\ \cline{3-12}

& & \multirow{3}{*}{\begin{tabular}{@{}c@{}} Microsoft Office \\(2019) \end{tabular}} 
& CVE-2024-38021 & 8.8 & 2.8 & 5.9 & 0.00091 & 0 & False & 0 & 46 \\ \cline{4-12} 
&  &  & CVE-2023-35311 & 8.8 & 2.8 & 5.9 & 0.01769 & 1 & False & 0 & 28 \\ \cline{4-12} 
&  &  & CVE-2021-38646 & 7.8 & 1.8 & 5.9 & 0.17921 & 1 & False & 1 & 27 \\ \cline{3-12}

& &  Google Chrome (122.0.6261.57) 
& CVE-2024-5839 & 8.8 & 2.8 & 5.9 & 0.00054 & 0 & False & 0 & 47 \\ \hline

\begin{tabular}{@{}c@{}} External \\Firewall \end{tabular}
& \multirow{1}{*}{2} 
& Cisco ASA 5505 (9.6(4)) & CVE-2020-3187 & 9.1 & 3.9 & 5.2 & 0.97297 & 0 & False & 0 & 13 \\ \hline

\multirow{2}{*}{\begin{tabular}{@{}c@{}} Internal \\Firewall \end{tabular}} 
& \multirow{2}{*}{2} 
& \multirow{2}{*}{\begin{tabular}{@{}c@{}} Netgate pfSense \\(2.7.0) \end{tabular}} 
& CVE-2023-42327 & 5.4 & 2.3 & 2.7 & 0.00083 & 0 & True & 0 & 36 \\ \cline{4-12}
&  &  & CVE-2023-48123 & 8.8 & 2.8 & 5.9 & 0.0017 & 0 & False & 0 & 29 \\ \hline

\end{tabular}}
\label{tab:completeRank_II}
\end{table*}

\end{document}